\documentclass[preprintnumbers,secnumarabic,twocolumn]{revtex4}

\usepackage{amsmath,amsthm,amssymb,amsfonts}
\usepackage{mathrsfs,bbm}
\usepackage{graphicx} % available on arxiv.org

\newcommand*{\smgroup}{\ensuremath{SU(3)_C \times SU(2)_L \times U(1)_Y}}
\newcommand*{\ewgroup}{\ensuremath{SU(2)_L \times U(1)_Y}}
\newcommand*{\clgroup}{\ensuremath{SU(3)_C}}
\newcommand*{\wigroup}{\ensuremath{SU(2)_L}}
\newcommand*{\hygroup}{\ensuremath{U(1)_Y}}
\newcommand*{\CP}{\ensuremath{\text{CP}}}
\newcommand*{\CPs}{\ensuremath{\mathrm{CP}_s}}
\newcommand*{\CPg}{\ensuremath{\text{CP}_g}}
\newcommand*{\CPi}{\ensuremath{\mathrm{CP}_g^{(i)}}}
\newcommand*{\CPii}{\ensuremath{\text{CP}_g^{(ii)}}}
\newcommand*{\CPa}{\ensuremath{\text{CP}_{g,1}^{(ii)}}}
\newcommand*{\CPb}{\ensuremath{\text{CP}_{g,2}^{(ii)}}}
\newcommand*{\CPc}{\ensuremath{\text{CP}_{g,3}^{(ii)}}}
\newcommand*{\unitmatrix}{\mathbbm{1}}
\newcommand*{\vev}[1]{\left\langle{#1}\right\rangle}
\newcommand*{\abs}[1]{\left\lvert {#1} \right\rvert} % abs. value or norm
\newcommand*{\twomat}[1]{\underline{#1}}             % 2x2 matrix
\newcommand*{\tvec}[1]{\ensuremath{\boldsymbol{\mathrm{#1}}}}           % 3 vec
\newcommand*{\tmat}[1]{#1}
\newcommand*{\fvec}[1]{\ensuremath{\boldsymbol{\mathrm{\tilde{#1}}}}  } % 4 vec
\newcommand*{\fmat}[1]{\tilde{#1}}                   % 4x4 matrix
\newcommand*{\trans}{\mathrm{T}}                     % transposed
\newcommand*{\by}{\!\times\!}                        % for nxn matrices
\DeclareMathOperator{\diag}{diag}

\begin{document}

\preprint{HD-THEP-07-30}

\title{A new type of \CP\ symmetry, family replication and fermion mass 
       hierarchies}

\author{M. Maniatis}
    \email[E-mail: ]{M.Maniatis@thphys.uni-heidelberg.de}
\author{A. von Manteuffel}
    \email[E-mail: ]{A.v.Manteuffel@thphys.uni-heidelberg.de}
\author{O. Nachtmann}
    \email[E-mail: ]{O.Nachtmann@thphys.uni-heidelberg.de}

\affiliation{
Institut f\"ur Theoretische Physik, Philosophenweg 16, 69120
Heidelberg, Germany
}

\begin{abstract}
We study a two-Higgs-doublet model with four generalised
\CP\ symmetries in the scalar sector.
Electroweak symmetry breaking leads automatically to spontaneous
breaking of two of them.
We require that these four \CP\ symmetries can be extended from
the scalar sector to the full Lagrangian and call this requirement the
principle of maximal \CP\ invariance.
The Yukawa interactions of the fermions are severely restricted by
this requirement.
In particular, a single fermion family cannot be coupled to the
Higgs fields.
For two fermion families, however, this is possible.
Enforcing the absence of flavour-changing neutral currents,
we find degenerate masses in both families
or one family massless and one massive.
In the latter case the Lagrangian is highly symmetric, with
the mass hierarchy being generated by electroweak symmetry
breaking.
Adding a third family uncoupled to the Higgs fields and thus
keeping it massless we get a model which gives a rough
approximation of some features of the fermions observed in Nature.
We discuss a number of predictions of the model which may be checked in future experiments at the LHC.
\end{abstract}

\maketitle

%%%%%%%%%%%%%%%%%%%%%%%%%%%%%%%%%%%%%%%%%%%%%%%%%%%%%%%%%%%%%%%%%%%%%%%%%%%%%%
\section{Introduction}
%%%%%%%%%%%%%%%%%%%%%%%%%%%%%%%%%%%%%%%%%%%%%%%%%%%%%%%%%%%%%%%%%%%%%%%%%%%%%%
\label{sec-introduction}

In the Standard Model of particle physics (SM) we have three families of 
fermions, leptons and quarks, the electroweak gauge bosons $\gamma$, $W^\pm$,
$Z$, the gluon $G$ as the strong interaction gauge boson and one scalar Higgs 
doublet field
\cite{Weinberg:1967tq, Salam:1968rm, Glashow:1970gm}.
For an introduction to the physics of the SM see for instance
\cite{Nachtmann:1990ta}.
Our notations and kinematics conventions follow this reference.
The SM gives, however, no explanation why there should be a replication of 
families in Nature.
Also, the fermion masses, arising from the Yukawa interactions of the 
fermions with the Higgs field, remain arbitrary.

In this paper we present some considerations based on a two-Higgs-doublet model 
(THDM).
We show that a certain type of \CP\ symmetry, which is possible for a THDM 
forbids a non-zero Yukawa coupling to {\em one} fermion family only.
However, an invariant coupling to {\em two} fermion families can be constructed.
Moreover, this new type of \CP\ symmetry is automatically spontaneously broken
by the electroweak symmetry breaking  (EWSB).
As we shall show, this also leads to interesting results for the masses in
the two fermion families.
In essence we find that in this type of theories absence of large 
flavour-changing neutral currents (FCNCs) can only be achieved if the corresponding
fermions in the two families have equal masses or if one family is massive,
one massless.

Various aspects of two-Higgs-doublet models have been investigated in the 
literature; see for instance 
\cite{Gunion:2005ja,
Nishi:2006tg,
Ginzburg:2004vp,
Ivanov:2006yq,
Ivanov:2007de,
Kobayashi:1973fv,
Barroso:2007rr,
Gerard:2007kn,
Barbieri:2005kf,
Fromme:2006cm,
Branco:2005em}
and references therein. 
In our group we have made a systematic study of the stability and symmetry breaking 
in the most general THDM \cite{Maniatis:2006fs}.
In \cite{Maniatis:2007vn} we have made a systematic investigation of all
possible types of generalised \CP\ transformations for the two Higgs doublets.
We have also given the criteria for \CP\ invariance of the two-Higgs-doublet 
potential and for spontaneous \CP\ violation.
In particular, we classified the generalised \CP\ transformations as type (i) 
and type (ii).
The structure of the Higgs sector of theories with type (i) \CP\ invariance was 
discussed.
In the present work we continue the investigation of THDMs with type (i) \CP\ 
invariance in view of the possibilities for invariant fermion--Higgs-boson 
couplings in this framework.
Indeed, the paper \cite{Maniatis:2007vn} and the present paper are companion papers. 
Therefore sections and equations of \cite{Maniatis:2007vn} will be quoted as section I.1 
etc. and (I.1), (I.2) etc. respectively.

Our paper is organised as follows.
In section~\ref{sec-model} we recall the main results of \cite{Maniatis:2007vn}
concerning THDMs with generalised \CP\ invariance of type~(i).
In section~\ref{sec-higgses} we discuss the vacuum expectation values (VEVs) 
and the physical Higgs mass spectrum of our models.
In section~\ref{sec-fermions} we introduce the fermion families and consider 
their coupling to the Higgs fields.
In section~\ref{sec-discussion} we collect our results and discuss their
physical consequences.
Section~\ref{sec-conclusions} contains our conclusions.
Detailed derivations of various results are presented in the appendices.

%%%%%%%%%%%%%%%%%%%%%%%%%%%%%%%%%%%%%%%%%%%%%%%%%%%%%%%%%%%%%%%%%%%%%%%%%%%%%%
\section{The THDM with \CP~type~($\mathrm{\mathbf{i}}$) invariance}
%%%%%%%%%%%%%%%%%%%%%%%%%%%%%%%%%%%%%%%%%%%%%%%%%%%%%%%%%%%%%%%%%%%%%%%%%%%%%%
\label{sec-model}

We consider models with the particle content as in the SM but with $n$ fermion 
families $(n=1,2,3)$ and with two Higgs doublet fields instead of one.
The two Higgs fields are denoted by 
\begin{equation}\label{2.1}
\varphi_i(x) = \begin{pmatrix} \varphi_i^+(x) \\ \varphi_i^0(x) \end{pmatrix}
\end{equation}
with $i=1,2$.
The gauge group of our models is assumed to be the usual one, \smgroup.
In the following the gauge group \clgroup\ of strong interactions will play no 
role.
We shall be concerned with the gauge groups of weak isospin, \wigroup, and weak 
hypercharge \hygroup.
Both Higgs doublets in \eqref{2.1} are assigned weak hypercharge $y=1/2$.
The most general gauge-invariant Lagrange density can then be written as
\begin{equation}\label{2.2}
\mathscr{L}_{\text{THDM}} = \mathscr{L}_\varphi + \mathscr{L}_\text{Yuk} + 
  \mathscr{L}_\text{FB}\,.
\end{equation}
Here $\mathscr{L}_\text{FB}$ is the standard gauge kinetic Lagrange density
for fermions and gauge bosons; see for instance~\cite{Nachtmann:1990ta}.
The Higgs-boson Lagrangian is 
\begin{equation}\label{2.3}
\mathscr{L}_\varphi = \sum_{i=1,2}
  \left( D_\mu \varphi_i \right)^\dagger \left( D^\mu \varphi_i \right)
  - V(\varphi_1,\varphi_2)\,,
\end{equation}
with $V(\varphi_1,\varphi_2)$ the Higgs-boson potential.
Finally, the Yukawa term, denoted by $\mathscr{L}_\text{Yuk}$, contains the 
Higgs-boson-fermion couplings which will be the main focus of study in this 
paper. 

Let us, however, first recall the main result from~\cite{Maniatis:2006fs} 
and~\cite{Maniatis:2007vn} 
concerning the Higgs potential $V$ and \CP\ transformations in the Higgs sector.
We use the framework of gauge-invariant functions.
For this we introduce the $2 \by 2$ matrix of the Higgs fields (see (A.2) of 
\cite{Maniatis:2006fs} and (I.8)),
\begin{equation}\label{2.3a}
\phi(x) = \begin{pmatrix}
  \varphi_1^+(x) & \varphi_1^0(x) \\
  \varphi_2^+(x) & \varphi_2^0(x) \end{pmatrix}\,.
\end{equation}
Similarly, the scalar products of two Higgs fields, 
$\varphi^\dagger_i(x)\varphi_j(x)$ $(i,j\in \{1,2\})$ are grouped into a
$2\by 2$ hermitian matrix
\begin{align}\label{2.4}
\twomat{K}(x) &= \phi(x)\phi^\dagger(x)\notag\\
  &= \frac{1}{2}\left( K_0(x)\unitmatrix_2
      + \tvec{K}(x)\,\tvec{\sigma} \right)\,.
\end{align}
Here we have expanded $\twomat{K}(x)$ in terms of the unit matrix 
$\unitmatrix_2$ and the Pauli matrices $\sigma^a\ (a=1,2,3)$, thus defining the 
gauge-invariant functions
\begin{equation}\label{2.5}
\fvec{K}(x) = \begin{pmatrix} K_0(x) \\ \tvec{K}(x) \end{pmatrix}\,.
\end{equation}
These form a real four-vector parametrising the gauge orbits of the Higgs 
fields.
As we see immediately from \eqref{2.4} $\twomat{K}(x)$ is a positive 
semidefinite $2 \by 2$ matrix.
This implies that $\fvec{K}(x)$ must be on or inside the forward light cone:
\begin{align}\label{2.6}
\fvec{K}^\trans(x)\, \fmat{g}\, \fvec{K}(x) & \ge 0\,,\notag\\
K_0(x) &\ge 0\,,
\end{align}
where 
\begin{equation}\label{2.7}
\fmat{g} = \diag(1,-1,-1,-1)\,.
\end{equation}
The most general \ewgroup\,-invariant potential can now be written as
\begin{align}\label{2.8}
V =& \fvec{\xi}^\trans\,\fvec{K}(x)
   + \fvec{K}^\trans(x)\, \fmat{E}\, \fvec{K}(x) \notag\\
 =& \xi_0\, K_0(x) + \tvec{\xi}^\trans\, \tvec{K}(x)
  + \eta_{00}\, K_0^2(x)\notag\\ 
  &+ 2\, K_0(x)\,\tvec{\eta}^\trans\, \tvec{K}(x)
  + \tvec{K}^\trans(x)\, \tmat{E}\, \tvec{K}(x)\,.
\end{align}
Here the potential parameters are a real four-vector and a real symmetric
$4\by 4$ matrix:
\begin{gather}
\fvec{\xi} = \begin{pmatrix}\xi_0 \\ \tvec{\xi} \end{pmatrix}\,,
\qquad
\fmat{E} = \begin{pmatrix}
  \eta_{00} & \tvec{\eta}^\trans \\
  \tvec{\eta} & \tmat{E} \end{pmatrix}\,,
\notag\\
\tvec{\xi} = \begin{pmatrix}\xi_1 \\ \xi_2 \\ \xi_3 \end{pmatrix}\,,
\qquad
\tvec{\eta} = \begin{pmatrix}\eta_1 \\ \eta_2 \\ \eta_3 \end{pmatrix}\,,\notag\\
\tmat{E} = \tmat{E}^\trans = \begin{pmatrix}
  \eta_{11} & \eta_{12} & \eta_{13} \\
  \eta_{21} & \eta_{22} & \eta_{23} \\
  \eta_{31} & \eta_{32} & \eta_{33} \end{pmatrix}.
\label{2.9}
\end{gather}

Under a basis change of the Higgs doublets
\begin{equation}\label{2.9a}
\varphi_i'(x) = U_{ij}\,\varphi_j(x)\,,
\qquad i,j \in \{1,2\}\,,
\end{equation}
with $U=(U_{ij})\in U(2)$ the functions $K_0(x)$, $\tvec K(x)$ transform as
\begin{align}\label{2.10}
K_0'(x) &= K_0(x)\,,\notag\\
\tvec{K}'(x) &= R(U)\,\tvec{K}(x)\,.
\end{align}
Here $R(U)$ is a proper rotation matrix $\big(R(U)\in SO(3)\big)$ obtained from 
\begin{equation}\label{2.11}
U^\dagger \sigma^a U = R_{ab}(U) \sigma^b\,,
\quad a,b \in \{1,2,3\}\,.
\end{equation}
The potential \eqref{2.8} stays the same under such a basis change if we 
transform the parameters \eqref{2.9} as follows
\begin{gather}
\begin{aligned}
\xi_0' &= \xi_0\,,\qquad         & \tvec{\xi}' = \tmat{R}(U)\,\tvec{\xi}\,,\\
\eta_{00}' &= \eta_{00}\,, & \tvec{\eta}' = \tmat{R}(U)\,\tvec{\eta}\,,
\end{aligned}
\notag\\
\label{2.12}
\tmat{E}' = \tmat{R}(U)\, \tmat{E}\, \tmat{R}^\trans(U)\,.
\end{gather}
This can for instance be used to diagonalise the real symmetric matrix 
$\tmat{E}$. 

The precise conditions for the potential \eqref{2.8} to be stable and to lead 
to the EWSB observed in Nature were spelled out in \cite{Maniatis:2006fs}.
A thorough investigation of the possibilities for \CP\ invariance and 
spontaneous \CP\ violation for the Higgs Lagrangian \eqref{2.3} with the 
potential \eqref{2.8} was done in \cite{Maniatis:2007vn}.
There, we gave a classification of possible generalised \CP\ transformations of 
the Lagrangian \eqref{2.3} in type (i) and (ii).
This classification is geometrically motivated:
The \CPi\ transformation of type~(i) corresponds to the point reflection
at the origin in $\tvec{K}$-space, whereas a type~(ii) \CPg\ transformation
corresponds to a reflection on a plane in $\tvec{K}$-space.
Note that it is not possible to convert type~(i) and type~(ii) transformations
into each other by a change of basis.

In terms of the fields, the type~(i) generalised \CP\ transformation is denoted
by \CPi\ and defined by (see (I.54))
\begin{alignat}{2}\label{2.13}
\CPi\,:\;
&&W^\mu(x) &\rightarrow -W_\mu^\trans(x')\,,\notag \\
&&B^\mu(x) &\rightarrow -B_\mu(x')\,,\\[2mm]
\label{2.13a}
&&\varphi_i(x) &\rightarrow \epsilon_{ij} \varphi_j^\ast(x')\,,\notag\\
&&\phi(x)      &\rightarrow \epsilon \phi^\ast(x')\,.
\end{alignat}
Here 
\begin{equation}\label{2.14}
x = \begin{pmatrix} x^0 \\ \tvec{x} \end{pmatrix}, \qquad
x' = \begin{pmatrix} x^0 \\ -\tvec{x} \end{pmatrix},
\end{equation}
and
\begin{equation}\label{2.14a}
W^\mu(x) = W^{\mu\,a}(x)\frac{\tau^a}{2}
\end{equation}
is the matrix of $W$ four-potentials with $\tau^a\ (a=1,2,3)$ the Pauli 
matrices.
Furthermore, $B^\mu(x)$ is the hypercharge four-potential and 
\begin{equation}\label{2.15}
\epsilon = \begin{pmatrix} \phantom{+}0 & \phantom{+}1 \\ -1 & \phantom{+}0 \end{pmatrix}\,.
\end{equation}
With \eqref{2.13a} we find for the gauge-invariant functions
\begin{alignat}{2}\label{2.16}
\CPi\,:\;
&&K_0(x) &\rightarrow K_0(x')\,,\notag\\
&&\tvec{K}(x) &\rightarrow -\tvec{K}(x')\,.
\end{alignat}
That is, the \CPi\ transformation \eqref{2.13a} corresponds to the point 
reflection at the origin in $\tvec K$-space in addition to the change of 
argument $x\to x'$.
Twofold application of the \CPi\ transformation gives the original Higgs fields
with a minus sign.
This minus sign corresponds to a hypercharge transformation,
and thus drops out when considering the twofold \CPi\ transformation~\eqref{2.16}
of the gauge-invariant functions.
Note that the \CPi\ transformation of the fields given by~(\ref{2.13a})
has the {\em same} form in {\em any} Higgs basis, up to gauge
transformations. See section I.3 and appendix I.B.

The type (ii) generalised \CP\ transformations are those which reduce to the 
standard \CP\ transformation in a suitable basis for the Higgs doublets
(see section I.3.2).
That is, after a suitable basis change \eqref{2.9a} we have
\begin{equation}\label{2.16a}
\CPii\,:\;
\varphi'_i(x) \rightarrow \varphi'_i{}^\ast(x'),\quad i\in \{1,2\}\,,
\end{equation}
whereas the transformation of the gauge potentials stays as in \eqref{2.13}. 
For the original Higgs basis we have then:
\begin{equation}\label{2.16b}
\CPii\,:\;
\varphi_i(x) \rightarrow (U^{-1}U^\ast)_{ij}\, \varphi_j{}^\ast(x')\,.
\end{equation}
Here, the argument change and complex conjugation of the Higgs fields 
$\varphi_i(x)$ is supplemented by a basis change.
For the gauge-invariant functions these type (ii) \CPg\ transformations 
\eqref{2.16b} correspond to reflections on a plane in $\tvec K$-space.
These reflections are orthogonally equivalent to $R_2$, the reflection on the
1--3 plane.
Indeed, let us define the reflections on the coordinate planes as
\begin{align}\label{2.17}
R_1 &= \diag(-1,\phantom{-}1,\phantom{-}1)\,,\notag\\
R_2 &= \diag(\phantom{-}1,-1,\phantom{-}1)\,,\notag\\
R_3 &= \diag(\phantom{-}1,\phantom{-}1,-1)\,.
\end{align}
Then we find from \eqref{2.16b}
\begin{alignat}{2}
\CPii\,:\qquad
&&K_0(x) &\rightarrow K_0(x)\,,\notag\\
&&\tvec{K}(x) &\rightarrow \bar{R}_\varphi\, \tvec{K}(x')\,,
\end{alignat}
where we have with $R(U)$ from \eqref{2.11}
\begin{equation}\label{2.18}
\bar{R}_\varphi = R^\trans(U)\, R_2\, R(U)\,.
\end{equation}
Note that a twofold \CPii\ transformation reproduces the original Higgs fields
without an additional phase.

Invariance of the potential \eqref{2.8} under a generalised \CP\ transformation
was found in section I.3 to require for the parameters \eqref{2.9}
\begin{equation}\label{2.19}
\bar{R}\, \tvec{\xi} = \tvec{\xi}\,,\quad
\bar{R}\, \tvec{\eta} = \tvec{\eta}\,,\quad
\bar{R}\, E\, \bar{R}^\trans = E\,.
\end{equation}
Here $\bar R$ is the improper rotation matrix in $\tvec{K}$-space corresponding 
to the generalised \CP\ symmetry.
A generalised \CPg\ symmetry of type~(i) corresponds to 
$\bar R=-\unitmatrix_3$ whereas type (ii) corresponds to
$\bar R=\bar R_\varphi$, \eqref{2.18}.

Now we can write down the most general potential having \CPi\ invariance; see 
theorem I.2.
We must have $\tvec{\xi}=0$ and $\tvec{\eta}=0$ in order to satisfy 
\eqref{2.19}.
Thus,
\begin{equation}\label{2.20}
V =  \xi_0 K_0(x) + \eta_{00} K_0^2(x) + \tvec{K}^\trans(x)\, \tmat{E}\, \tvec{K}(x)\,.
\end{equation}
In the following we shall always suppose that by a basis change as in 
\eqref{2.9a} and \eqref{2.12} we have diagonalised $E$ and ordered the eigenvalues 
as follows:
\begin{equation}\label{2.21}
E=\diag(\mu_1,\mu_2,\mu_3)\,,\quad \mu_1 \geq \mu_2 \geq \mu_3\,.
\end{equation}

In theorem I.5 we found that the potential \eqref{2.20} leads to a 
stable theory with the correct EWSB and no zero mass charged Higgs bosons
if and only if
\begin{align}\label{2.22}
\eta_{00} &> 0,\notag\\
\mu_a + \eta_{00} &> 0\quad\text{for } a=1,2,3\,,\notag\\
\xi_0 &< 0\,,\notag\\
\mu_3 &< 0\,.
\end{align}
In the following we shall always suppose these conditions to hold.

%%%%%%%%%%%%%%%%%%%%%%%%%%%%%%%%%%%%%%%%%%%%%%%%%%%%%%%%%%%%%%%%%%%%%%%%%%%%%%
\section{The vacuum expectation values and the~Higgs~mass~spectrum}
%%%%%%%%%%%%%%%%%%%%%%%%%%%%%%%%%%%%%%%%%%%%%%%%%%%%%%%%%%%%%%%%%%%%%%%%%%%%%%
\label{sec-higgses}

The vacuum solution for the Higgs fields, denoted by
\begin{equation}\label{3.1}
\vev{\varphi_j}:=\vev{\varphi_j(x)} =
  \begin{pmatrix} v_j^{+} \\ v_j^0 \end{pmatrix}\,,\qquad
  j=1,2
\end{equation}
is obtained from the global minimum of $V$ \eqref{2.20}.
The corresponding matrices $\vev{\phi}$ and $\twomat{K}$ are 
\begin{equation}\label{3.2}
\vev{\phi}:=\vev{\phi(x)}=
  \begin{pmatrix} v_1^+ & v_1^0 \\ v_2^+ & v_2^0 \end{pmatrix}\,,
\end{equation}
\begin{equation}\label{3.3}
\twomat{K} = \vev{\phi}\vev{\phi}^\dagger
  = \frac{1}{2}(K_0\,\unitmatrix_2 + \tvec{K}\,\tvec{\sigma}),
\end{equation}
\begin{equation}\label{3.4}
\fvec{K} = \begin{pmatrix} K_0 \\ \tvec{K} \end{pmatrix}\,.
\end{equation}
In appendix I.B we have already discussed this vacuum solution.
We get (see (I.B.41))
\begin{equation}\label{3.5}
\fvec{K} = \begin{pmatrix} K^0 \\ \tvec{K} \end{pmatrix}
  = \frac{-\xi_0}{2 (\eta_{00} + \mu_3)}
     \begin{pmatrix} 1\\ 0\\ 0\\ 1 \end{pmatrix}\,.
\end{equation}

In section 7 of \cite{Maniatis:2006fs} a general discussion of the structure of the Higgs 
sector in THDMs after EWSB was given.
The basis choice in this reference coincides with our present choice.
We can, therefore, set for the Higgs field vacuum expectation values
\begin{align}\label{3.6}
\vev{\varphi_1} &= \frac{1}{\sqrt{2}}\begin{pmatrix} 0 \\ v_0 \end{pmatrix}\,,\notag\\
\vev{\varphi_2} &= \begin{pmatrix} 0 \\ 0 \end{pmatrix}\,,\notag\\
\vev{\phi} &=
  \frac{1}{\sqrt{2}} \begin{pmatrix} 0 & v_0 \\ 0 & 0 \end{pmatrix}\,.
\end{align}
Here
\begin{equation}\label{3.7}
v_0 \approx 246\text{ GeV}
\end{equation}
is the standard Higgs vacuum expectation value (see for instance~\cite{Nachtmann:1990ta}). 
Furthermore, we use the unitary gauge and introduce the shifted, that is, the 
physical Higgs fields as in (129)~ff of \cite{Maniatis:2006fs}.
This leads to 
\begin{align}
\label{3.7a}
\varphi_1(x) &=
  \frac{1}{\sqrt{2}}\begin{pmatrix} 0 \\ v_0 + \rho'(x) \end{pmatrix}\,,\\
\label{3.8}
\varphi_2(x) &=
  \begin{pmatrix} H^+(x)\\ \frac{1}{\sqrt{2}}( h'(x) + i\,h^{''}(x))
  \end{pmatrix}\,.
\end{align}
Here $\rho'(x)$, $h'(x)$ and $h''(x)$ are the three real fields corresponding
to the physical neutral Higgs particles and $H^+(x)$ is the complex field 
corresponding to the physical charged Higgs particle.
We set 
\begin{equation}\label{3.9}
H^-(x) = (H^+(x))^\ast\,.
\end{equation}
From the results of section 7 of \cite{Maniatis:2006fs} and appendix I.B
we can now immediately read off a number of relations. 

The Lagrange multiplier $u_0$ corresponding to the global minimum of the 
potential $V$ \eqref{2.20} is
\begin{equation}\label{3.10}
u_0 = -\mu_3\,.
\end{equation}
Inserting \eqref{3.6} in \eqref{3.3} we get for the vacuum four-vector
$\fvec{K}$ (see (134) of \cite{Maniatis:2006fs})
\begin{equation}\label{3.11}
\fvec{K} = \frac{1}{2}v_0^2 \begin{pmatrix} 1\\0\\0\\1 \end{pmatrix}\,.
\end{equation}
Comparison with \eqref{3.5} gives
\begin{align}\label{3.12}
v_0^2 &= \frac{-\xi_0}{\eta_{00} + \mu_3} =
  \frac{\abs{\xi_0}}{\eta_{00}-\abs{\mu_3}}\,,\notag\\
v_0 &= \sqrt[\leftroot{3}\uproot{-30}+]{\frac{-\xi_0}{\eta_{00}+\mu_3}}\,.
\end{align}

The mass squared of the charged Higgs particles is, according to (145) of 
\cite{Maniatis:2006fs},
\begin{align}\label{3.13}
m_{H^\pm}^2 &= 2 u_0\, v_0^2 = 2(-\mu_3)v_0^2\notag\\
  &= \frac{2 \mu_3 \xi_0}{\eta_{00}+\mu_3}\,.
\end{align}
The mass matrix squared of the neutral Higgs particles is obtained from (144) 
of \cite{Maniatis:2006fs} as follows:
\begin{equation}\label{3.14}
\mathcal{M}_{\mathrm{neutral}}^2 =
  2 \begin{pmatrix}
		-\xi_0 & 0 & 0\\
		0 & v_0^2(\mu_1-\mu_3) & 0\\
		0 & 0 & v_0^2(\mu_2-\mu_3)
  \end{pmatrix}
\end{equation}
with the ordering $(\rho',h',h'')$ for the fields.
We see that $\mathcal{M}_{\mathrm{neutral}}^2$ is already diagonal with our 
choice of basis. 
Thus we have
\begin{align}\label{3.15}
m_{\rho'}^2 &= 2(-\xi_0)\,,\notag\\
m_{h'}^2    &= 2\,v_0^2\,(\mu_1-\mu_3)\,,\notag\\
m_{h''}^2   &= 2\,v_0^2\,(\mu_2-\mu_3)\,.
\end{align}
In the following we shall require that none of the neutral physical Higgs 
particles is massless and that there is no mass degeneracy between $h'$ and 
$h''$.
This implies from \eqref{3.15} the condition
\begin{equation}\label{3.16}
\mu_1 > \mu_2 > \mu_3
\end{equation}
which is slightly stricter than \eqref{2.21}.

Our Higgs potential \eqref{2.20} has five parameters $\xi_0$, $\eta_{00}$,
$\mu_1$, $\mu_2$, $\mu_3$.
We can now express these in terms of the five physical quantities $v_0^2$,
$m_{H^\pm}^2$, $m_{\rho'}^2$, $m_{h'}^2$, $m_{h''}^2$.
This gives
\begin{align}\label{3.16a}
\xi_0     &= -\frac{1}{2} m_{\rho'}^2\,,\notag\\
\eta_{00} &= \frac{1}{2 v_0^2} \left( m_{H^\pm}^2 + m_{\rho'}^2 \right)\,,\notag\\
\mu_1     &= \frac{1}{2 v_0^2} \left( m_{h'}^2    - m_{H^\pm}^2 \right)\,,\notag\\
\mu_2     &= \frac{1}{2 v_0^2} \left( m_{h''}^2   - m_{H^\pm}^2 \right)\,,\notag\\
\mu_3     &= -\frac{1}{2 v_0^2} m_{H^\pm}^2\,.
\end{align}
For positive squared masses and $m_{h'}^2 > m_{h''}^2$ the conditions
\eqref{2.22} and \eqref{3.16} are always satisfied.

Let us next discuss the \CP\ symmetries of our model and the \CP\ 
transformation properties of the vacuum expectation values and of the
physical fields.
The Higgs Lagrangian \eqref{2.3} with the potential \eqref{2.20} for which we
require \eqref{3.16} to hold, allows for exactly four generalised \CP\  
transformations, \CPi, \CPa, \CPb\ and \CPc\, as defined below.
In all cases the gauge potentials are transformed according to \eqref{2.13}. 
But the transformation of the Higgs fields and of the gauge-invariant 
functions $K_0(x)$, $\tvec{K}(x)$ is different.

Our basic type $(i)$ \CP\ transformation, \CPi, transforms the Higgs 
fields and the gauge-invariant functions according to \eqref{2.13a} and 
\eqref{2.16}, respectively.

For the type (ii) transformation \CPa\ we set
\begin{alignat}{2}\label{3.17}
\CPa\,:\quad
&&\varphi_i(x) &\rightarrow \sigma^3_{ij}\varphi_j^\ast(x')\,,\notag\\
&&\varphi_1(x) &\rightarrow \phantom{+}\varphi_1^\ast(x')\,,\notag\\
&&\varphi_2(x) &\rightarrow -\varphi_2^\ast(x')\,.
\end{alignat}
This implies
\begin{alignat}{2}\label{3.18}
\CPa\,:\quad
&&K_0(x)      &\rightarrow K_0(x')\,,\notag\\
&&\tvec{K}(x) &\rightarrow R_1\,\tvec{K}(x')\,,
\end{alignat}
with $R_1$ the matrix of the reflection on the 2--3 plane; see \eqref{2.17}.

The type (ii) transformation \CPb\ is the standard CP 
transformation, $\CPs$, for the Higgs fields, where
\begin{alignat}{2}\label{3.19}
\CPb\,:\quad
&&\varphi_1(x) &\rightarrow \varphi_1^\ast(x')\,,\notag\\
&&\varphi_2(x) &\rightarrow \varphi_2^\ast(x')\,.
\end{alignat}
Here we get
\begin{alignat}{2}\label{3.20}
\CPb\,:\quad
&&K_0(x)      &\rightarrow K_0(x')\,,\notag\\
&&\tvec{K}(x) &\rightarrow R_2\,\tvec{K}(x')\,,
\end{alignat}
with $R_2$ the matrix of the reflection on the 1--3 plane; see \eqref{2.17}.

Finally, the transformation \CPc\ is defined by
\begin{alignat}{2}\label{3.20a}
\CPc\,:\quad
&&\varphi_i(x) &\rightarrow \sigma^1_{ij}\varphi_j^\ast(x')\,,\notag\\
&&\varphi_1(x) &\rightarrow \varphi_2^\ast(x')\,,\notag\\
&&\varphi_2(x) &\rightarrow \varphi_1^\ast(x')\,.
\end{alignat}
This implies
\begin{alignat}{2}\label{3.20b}
\CPc\,:\quad
&&K_0(x)      &\rightarrow K_0(x')\,,\notag\\
&&\tvec{K}(x) &\rightarrow R_3\,\tvec{K}(x')\,,
\end{alignat}
with $R_3$ the reflection on the 1--2 plane; see \eqref{2.17}.

Now we summarise the four different \CPg\ transformations for the 
Higgs fields as 
\begin{equation}\label{4.16}
\CPg\,:\quad
\varphi_i(x) \rightarrow W_{ij}\, \varphi_j^\ast(x')\,.
\end{equation}
The matrices $W=(W_{ij})$ for the various \CPg\ transformations are listed in 
Tab.~\ref{tab-cphiggs}; see \eqref{2.13a}, \eqref{3.17}, \eqref{3.19} and
\eqref{3.20a}.
We note that we could supplement an additional global phase factor on
the right-hand side of \eqref{4.16}.
However, such a global phase factor in the Higgs \CP\ transformation
drops out in the Higgs potential, and for the Yukawa terms
it may always be absorbed by proper redefinitions of the fermion fields,
as will be explained in the next section.
Therefore we may without loss of generality set this global phase factor to $1$.

\medskip\noindent
\begin{table}
\begin{tabular}{|c|c|}
\hline
\CPg & $W$\\
\hline\hline
\CPi &$\epsilon$\\
\CPa &$\sigma^3$\\
\CPb &${\mathbbm 1}_2$\\
\CPc &$\sigma^1$\\
\hline
\end{tabular}
\caption{\label{tab-cphiggs}
The matrices $W$ \eqref{4.16} for the four \CPg\ transformations.
}
\end{table}

Note that the symmetries \CPi, \CPa, \CPb\ and \CPc\ are not
independent since we have at the level of the transformation of the Higgs fields
the relation
\begin{equation}\label{3.20c}
\CPc = \CPa \circ \CPb \circ \CPi \,.
\end{equation}

Any of the above \CP\ symmetries is conserved by the vacuum if and only if the 
vacuum value $\tvec{K}$ satisfies 
\begin{equation}\label{3.21}
\bar{R}\, \tvec{K} = \tvec{K}\,.
\end{equation}
Here we have to insert $\bar R=-\unitmatrix_3$, $R_1$, $R_2$ and $R_3$ for the 
symmetries \CPi, \CPa\, \CPb\ and \CPc, respectively. 
Looking at the vacuum solution \eqref{3.5} for $\fvec{K}$ we see immediately 
that 
\begin{align}\label{3.22}
(-\unitmatrix_3) \tvec{K} &\neq \tvec{K}\,,\notag\\
R_1 \tvec{K} &= \tvec{K}\,,\notag\\
R_2 \tvec{K} &= \tvec{K}\,,\notag\\
R_3 \tvec{K} &\neq \tvec{K}\,.
\end{align}
Thus, the symmetries \CPi\ and \CPc\ are spontaneously broken, as we already 
know from theorem I.4.
On the other hand, the symmetries \CPa\ and \CPb\ are 
conserved by the vacuum.

Now we come to the \CP\ transformation properties of the physical Higgs fields 
defined in \eqref{3.7a} and \eqref{3.8}.
Under \CPi , which transforms the Higgs doublets according to 
\eqref{2.13a}, the physical Higgs fields have no definite transformation 
property.
This is alright, since \CPi\ is spontaneously broken.
For the unbroken symmetry \CPa\ we get from \eqref{3.7a}, 
\eqref{3.8} and \eqref{3.17}
\begin{alignat}{2}\label{3.23}
\CPa\,:\qquad
&&\rho'(x) &\rightarrow  \phantom{+}\rho'(x')\,,\notag\\
&&h'(x)    &\rightarrow  -h'(x')\,,\notag\\
&&h''(x)   &\rightarrow  \phantom{+}h''(x')\,,\notag\\
&&H^+(x)   &\rightarrow  -H^-(x')\,.
\end{alignat}
On the other hand, we obtain from \eqref{3.7a}, \eqref{3.8} and \eqref{3.19} for 
the \CPb\ symmetry
\begin{alignat}{2}\label{3.24}
\CPb\,:\qquad
&&\rho'(x) &\rightarrow  \phantom{+}\rho'(x')\,,\notag\\
&&h'(x)    &\rightarrow  \phantom{+}h'(x')\,,\notag\\
&&h''(x)   &\rightarrow  -h''(x')\,,\notag\\
&&H^+(x)   &\rightarrow  \phantom{+}H^-(x')\,.
\end{alignat}
We see that the field $\rho'$ is \CPa\ and \CPb\ even,
$h'$ is \CPa\ odd and $h''$ is \CPa\ even.
This role of $h'$ and $h''$ is interchanged for the symmetry \CPb;
see \eqref{3.24}.
We note, however, that this assignment of \CPii\ quantum numbers is to some
extent a convention, since we could have inserted extra global factors of $(-1)$
in \eqref{3.17} and \eqref{3.19}.
This would not change the transformation properties of the gauge-invariant
functions in \eqref{3.18} and \eqref{3.20} and thus have no physical 
consequence.

%%%%%%%%%%%%%%%%%%%%%%%%%%%%%%%%%%%%%%%%%%%%%%%%%%%%%%%%%%%%%%%%%%%%%%%%%%%%%%
\section{Fermions and their couplings to~the~Higgs~fields}
%%%%%%%%%%%%%%%%%%%%%%%%%%%%%%%%%%%%%%%%%%%%%%%%%%%%%%%%%%%%%%%%%%%%%%%%%%%%%%
\label{sec-fermions}

In this section we shall discuss the fermion families and their coupling to
the Higgs fields.
We shall require that the Higgs-fermion coupling, that is the Yukawa term
$\mathscr{L}_{\mathrm{Yuk}}$ in \eqref{2.2}, respects all four generalised
\CP\ symmetries, \CPi, \CPa, \CPb\ and \CPc, as introduced in
section~\ref{sec-higgses}.
We shall call this the ``principle of maximal \CP\ invariance''.
We shall show that this principle leads to interesting consequences.

Let us first introduce our notation for the fermions; see
Tab.~\ref{tab-fermions}.
We give the fermions a family index $j$ $(j=1,2,3)$ for ease of notation.
Thus, we set $l_1\equiv e$, $l_2\equiv \mu$ and $l_3\equiv\tau$ for the
leptons, $u_1\equiv u$, $u_2\equiv c$, $u_3\equiv t$ for the up type quarks
and $d_1\equiv d$, $d_2\equiv s$ and $d_3\equiv b$ for the down type quarks.
The indices $L$ and $R$ stand for left- and right-handed fields, respectively.
In Tab.~\ref{tab-fermions} we list also right-handed neutrinos.
The finding of neutrino oscillations suggests that they also form part of 
Nature.
In the following, however, we shall, as an approximation, consider the 
neutrinos as massless and ignore the $\nu_{j\,R}$ fields.
\begin{table}
\begin{tabular}{|c|c|c|}
\hline
Fermion & weak isospin $t$ & weak hypercharge $y$\\\hline\hline
$\begin{pmatrix}\nu_{jL}\\ l_{jL}\end{pmatrix}$&
$1/2$&
$-1/2$\\\hline
$\nu_{jR}$&$0$&$0$\\\hline
$l_{jR}$&$0$&$-1$\\\hline
$\begin{pmatrix}u_{jL}\\ d_{jL}\end{pmatrix}$&
$1/2$&$1/6$\\\hline
$u_{jR}$&$0$&$2/3$\\\hline
$d_{jR}$&$0$&$-1/3$\\\hline
\end{tabular}
\caption{\label{tab-fermions}
The fermion families, $j=1,2,3$, and their quantum numbers of weak 
isospin $t$ and weak hypercharge $y$.
}
\end{table}

%%%%%%%%%%%%%%%%%%%%%%%%%%%%%%%%%%%%%%%%%%%%%%%%%%%%%%%%%%%%%%%%%%%%%%%%%%%%%%
\subsection{The case of one family}
\label{sec-onefamily}

Now we study if we can couple one fermion family to the Higgs doublets in a 
\CPi-invariant way.
The most general $SU(2)_L \times U(1)_Y$ invariant Yukawa interaction for this 
case has the form (see chapter 22.4 of \cite{Nachtmann:1990ta} for the
analogous discussion in the framework of the SM)
\begin{align}\label{4.1}
\mathscr{L}_{\mathrm{Yuk}}(x) &=
  -\bar{l}_{1\,R}(x)\, c_{l,i}\, \varphi_i^\dagger(x)
     \begin{pmatrix} \nu_{1\,L}(x) \\ l_{1\,L}(x) \end{pmatrix}
\notag\\
  &\quad+\bar{u}_{1\,R}(x)\, c'_{q,i}\, \varphi_i^\trans(x) \epsilon
    \begin{pmatrix} u_{1\,L}(x) \\ d_{1\,L}(x) \end{pmatrix}
\notag\\
  &\quad
  -\bar{d}_{1\,R}(x)\, c_{q,i}\, \varphi_i^\dagger(x)
    \begin{pmatrix} u_{1\,L}(x) \\ d_{1\,L}(x) \end{pmatrix}
\notag\\
  &\quad
	+ h.c.
\end{align}
Here $c_{l,i}$, $c'_{q,i}$ and $c_{q,i}$ $(i=1,2)$ are arbitrary complex 
numbers.

Now we make a general ansatz for the \CPi\ transformation of the fermion
fields as follows:
\begin{align}\label{4.2} 
\begin{pmatrix} \nu_{1\,L}(x) \\ l_{1\,L}(x) \end{pmatrix}
	&\rightarrow e^{i\xi_1} \gamma^0\, S(C)\,
    \begin{pmatrix} \bar{\nu}_{1\,L}^\trans(x')\\ 
                    \bar{l}_{1\,L}^\trans(x')\end{pmatrix}\,,
\notag\\
l_{1\,R}(x)
	&\rightarrow e^{i\xi_2} \gamma^0\, S(C)\,
    \bar{l}_{1\,R}^\trans(x')\,,
\notag\\
\begin{pmatrix} u_{1\,L}(x) \\ d_{1\,L}(x) \end{pmatrix}
	&\rightarrow e^{i\xi_3} \gamma^0\, S(C)\,
    \begin{pmatrix} \bar{u}_{1\,L}^\trans(x')\\ 
                    \bar{d}_{1\,L}^\trans(x')\end{pmatrix}\,,
\notag\\
u_{1\,R}(x)
	&\rightarrow e^{i\xi_4} \gamma^0\, S(C)\,
    \bar{u}_{1\,R}^\trans(x')\,,
\notag\\
d_{1\,R}(x)
	&\rightarrow e^{i\xi_5} \gamma^0\, S(C)\,
    \bar{d}_{1\,R}^\trans(x')\,.
\end{align}
Here $x$, $x'$ are as in \eqref{2.14} and $\gamma^0$ and
$S(C):=i\gamma^2\gamma^0$ are the usual Dirac matrices for the parity and
charge conjugation transformations, respectively (see for instance chapter~4
of \cite{Nachtmann:1990ta}).
For generality we have inserted in \eqref{4.2}
arbitrary phase factors $e^{i\xi_a}$ with $\xi_a$ $(a=1,\ldots,5)$ real.
With \eqref{2.13a} and \eqref{4.2} we find the following
transformation of $\mathscr{L}_{\mathrm{Yuk}}(x)$, \eqref{4.1}:
\begin{align}\label{4.4}
&\mathscr{L}_{\mathrm{Yuk}}(x) \rightarrow 
\notag\\
& -e^{i(\xi_1-\xi_2)} \times\notag\\
&\quad   \begin{pmatrix} \bar{\nu}_{1\,L}(x'), & \bar{l}_{1\,L}(x')\end{pmatrix}
		c_{l,j}\, \epsilon_{ji}\, \varphi_i(x')\, l_{1\,R}(x')
\notag\\
&+e^{i(\xi_3-\xi_4)} \times\notag\\
&\quad   \begin{pmatrix} \bar{u}_{1\,L}(x'), & \bar{d}_{1\,L}(x')\end{pmatrix}
     c'_{q,j}\, \epsilon_{ji}\, \epsilon^\trans\, \varphi_i^\ast(x')\,
     u_{1\,R}(x')
\notag\\
&-e^{i(\xi_3-\xi_5)}\times\notag\\
&\quad   \begin{pmatrix} \bar{u}_{1\,L}(x'), & \bar{d}_{1\,L}(x')\end{pmatrix}
		c_{q,j}\, \epsilon_{ji}\, \varphi_i(x')\, d_{1\,R}(x')
\notag\\
&+ h.c.
\end{align}
Note that a possible additional global phase factor in the \CP\ transformation
of the Higgs fields, that is, on the right-hand side of \eqref{2.13a}, can be
absorbed by a redefinition of the phases $\xi_2$, $\xi_4$ and $\xi_5$ for the
right-handed fermions.
Similar remarks apply to the case of more than one fermion family.
Comparing \eqref{4.4} with \eqref{4.1} we see that we have \CPi\ invariance,
$\mathscr{L}_{\mathrm{Yuk}}(x) \to \mathscr{L}_{\mathrm{Yuk}}(x')$,
if and only if
\begin{align}\label{4.5}
c_{l,i}^\ast &= e^{i(\xi_1-\xi_2)} c_{l,j} \epsilon_{ji}\,,
\notag\\
c_{q,i}^{'\,\ast} &= e^{i(\xi_3-\xi_4)} c'_{q,j} \epsilon_{ji}\,,
\notag\\
c_{q,i}^\ast &= e^{i(\xi_3-\xi_5)} c_{q,j} \epsilon_{ji}
\end{align}
for $i=1,2$.
Explicitly we find from \eqref{4.5} for $c_{l,i}$:
\begin{align}\label{4.6}
c_{l,1}^\ast &=           -e^{i(\xi_1-\xi_2)}c_{l,2}\,,
\notag\\
c_{l,2}^\ast &= \phantom{-}e^{i(\xi_1-\xi_2)}c_{l,1}\,,
\end{align}
which has only the trivial solution
\begin{equation}\label{4.7}
c_{l,1} = c_{l,2} = 0\,.
\end{equation}
The same result is found for $c_{q,i}$ and $c'_{q,i}$.

We summarise these findings as follows.
A single fermion family (see Tab.~\ref{tab-fermions}) cannot be coupled
to the two-Higgs-doublet fields in a \CPi-symmetric way.
In other words: requiring \CPi\ symmetry for the Yukawa Lagrangian \eqref{4.1}
leads to
\begin{equation}\label{4.8}
c_{l,i}=c_{q,i}=c'_{q,i}=0\,,\qquad i=1,2\,,
\end{equation}
that is, to $\mathscr{L}_{\mathrm{Yuk}} \equiv 0$.

Our principle of maximal \CP\ invariance has as part of its requirements \CPi\ 
symmetry.
Thus, we have shown that a \emph{single} fermion family cannot be coupled
to the Higgs doublets in a way respecting our principle.

%%%%%%%%%%%%%%%%%%%%%%%%%%%%%%%%%%%%%%%%%%%%%%%%%%%%%%%%%%%%%%%%%%%%%%%%%%%%%%
\subsection{The case of two families, generalities}
\label{familiesgeneralities}

In this section we shall treat the case of two families where, for 
definiteness, we consider the families 2 and 3.
The most general Yukawa interaction of these families with the Higgs doublets 
can be written as
\begin{align}\label{4.9}
\mathscr{L}_{\mathrm{Yuk}}(x) &=
  -\bar{l}_{\alpha\,R}(x)\, C^{(j)}_{l\,\alpha\beta}\, \varphi_j^\dagger(x)
     \begin{pmatrix} \nu_{\beta\,L}(x) \\ l_{\beta\,L}(x) \end{pmatrix}
\notag\\
  &\quad
  +\bar{u}_{\alpha\,R}(x)\, C'^{(j)}_{q\,\alpha\beta}\, \varphi_j^\trans(x) 
    \,\epsilon
    \begin{pmatrix} u_{\beta\,L}(x) \\ d'_{\beta\,L}(x) \end{pmatrix}
\notag\\
  &\quad
  -\bar{d}'_{\alpha\,R}(x)\, C^{(j)}_{q\,\alpha\beta}\, \varphi_j^\dagger(x)
    \begin{pmatrix} u_{\beta\,L}(x) \\ d'_{\beta\,L}(x) \end{pmatrix}
\notag\\
  &\quad
	+ h.c.
\end{align}
Here $\alpha,\beta\in\{2,3\}$ are the family indices and $j\in\{1,2\}$ number 
the Higgs doublets.
The summation convention is used if not stated otherwise.
The $2\by 2$ matrices
$C^{(j)}_l   = ( C^{(j)}_{l\,\alpha\beta} )$, 
$C^{(j)}_q   = ( C^{(j)}_{q\,\alpha\beta} )$, 
$C'^{(j)}_q  = ( C'^{(j)}_{q\,\alpha\beta} )$
have, to start with, arbitrary complex entries.

Without changing the physical content of the theory we can make 
$U(2)$-rotations of the right-handed fields $l_{\alpha R}$,
$u_{\alpha R}$, $d'_{\alpha R}$ and the left-handed doublet fields
$(\nu_{\alpha L},~l_{\alpha L})^\trans$ and
$(u_{\alpha L},~d'_{\alpha L})^\trans$.
As in the SM we can use this to require, without loss of generality, for the 
matrices $C^{(1)}_l$, $C'^{(1)}_q$ and $C^{(1)}_q$ certain standard forms:
\begin{align}
\label{4.11}
  C^{(1)}_l &=
  \begin{pmatrix} c^{(1)}_{l\,2} & 0 \\ 0 & c^{(1)}_{l\,3} \end{pmatrix}\,,
  &\quad
	&c^{(1)}_{l\,2} \geq 0\,,\;\; c^{(1)}_{l\,3} \geq 0\,;
\\
\label{4.12}
  C'^{(1)}_q &=
  \begin{pmatrix} c^{(1)}_{u\,2} & 0 \\ 0 & c^{(1)}_{u\,3} \end{pmatrix}\,,
	&\quad
	&c^{(1)}_{u\,2} \geq 0\,,\;\; c^{(1)}_{u\,3} \geq 0\,;
\\
\label{4.13}
  C^{(1)}_q  &=
	V
  \begin{pmatrix} c^{(1)}_{d\,2} & 0 \\ 0 & c^{(1)}_{d\,3} \end{pmatrix}
	V^\dagger\,,
  &\quad
	&c^{(1)}_{d\,2} \geq 0\,,\;\; c^{(1)}_{d\,3} \geq 0\,,
\\
\label{4.14}
  V &=
    \begin{pmatrix} \phantom{+}\cos\vartheta & \sin\vartheta \\
                   -\sin\vartheta & \cos\vartheta \end{pmatrix}\,,
  &\quad
  &0 \leq \vartheta \leq \pi/2\,.
\end{align}
For the derivation of the corresponding results in the SM see, for instance, 
chapter 22.4 of \cite{Nachtmann:1990ta}.
The matrix $V=(V_{\alpha\beta})$ in \eqref{4.14} will turn out to be the 
Cabibbo--Kobayashi--Maskawa (CKM) matrix in the 2--3 sector.
As we shall see, in the basis of the fermion fields defined by 
\eqref{4.11}-\eqref{4.14} the fields $l_{\alpha\,R}$, $l_{\alpha\,L}$ and
$u_{\alpha\,R}$, $u_{\alpha\,L}$ correspond to mass eigenfields.
For the $d'$-fields defined in this basis the mass eigenstates will be
\begin{equation}\label{4.15}
d_{\alpha\,R,L}(x) = V^\dagger_{\alpha\beta}\,d'_{\beta\,R,L}(x)\,.
\end{equation}

In the following we shall always work in the fermion basis defined by 
\eqref{4.11}-\eqref{4.14} if not stated otherwise.

For the \CPg\ transformations of the fermions we make the generic ansatz:
\begin{alignat}{2}
\CPg\,\!\!:\!\!
\label{4.17}
\notag
&&  \begin{pmatrix} \nu_{\alpha\,L}(x) \\ l_{\alpha\,L}(x) \end{pmatrix}
  	&\rightarrow U^{(l)}_{L\,\alpha\beta}\,\gamma^0\, S(C)\,
      \begin{pmatrix} \bar{\nu}_{\beta\,L}^\trans(x') \\
                    \bar{l}_{\beta\,L}^\trans(x')\end{pmatrix}\,,
\notag  \\
&&  l_{\alpha\,R}(x) &\rightarrow
    U^{(l)}_{R\,\alpha\beta}\,\gamma^0\,S(C)\,\bar{l}_{\beta\,R}^\trans(x')\,,
\notag\\
&&  \begin{pmatrix} u_{\alpha\,L}(x) \\ d'_{\alpha\,L}(x) \end{pmatrix}
	  &\rightarrow U^{(u)}_{L\,\alpha\beta}\, \gamma^0\, S(C)\,
    \begin{pmatrix} \bar{u}^\trans_{\beta\,L}(x')\\ 
                    \bar{d}'^\trans_{\beta\,L}(x')\end{pmatrix}\,,
\notag\\
&&  u_{\alpha\,R}(x)
  	&\rightarrow U^{(u)}_{R\,\alpha\beta}\, \gamma^0\, S(C)\,
      \bar{u}_{\beta\,R}^\trans(x')\,,
\notag\\
&&  d'_{\alpha\,R}(x)
    &\rightarrow U^{(d)}_{R\,\alpha\beta}\, \gamma^0\, S(C)\,
      \bar{d}'^\trans_{\beta\,R}(x')\,.
\end{alignat}
All matrices
$U^{(l)}_L=(U^{(l)}_{L\,\alpha\beta})$ , \dots,
$U^{(d)}_R=(U^{(d)}_{R\,\alpha\beta})$
are supposed to be unitary
\begin{equation}\label{4.19}
U_L^{(l)}\,U_L^{(l)\,\dagger} = \dots =
U_R^{(d)}\,U_R^{(d)\,\dagger}
= \unitmatrix_2\,.
\end{equation}
Of course, these matrices $U^{(l)}_L$ etc.
will, in general, all be different for the four \CPg\ transformations which we 
consider.

Now we shall require that a \CPg\ transformation applied twice gives the 
original fields up to a phase factor.
Writing generically for any of the transformations \eqref{4.17}
%,\eqref{4.18} 
\begin{equation}\label{4.20}
\CPg\,:\;
\psi_\alpha(x) \rightarrow
   U_{\alpha\beta}\,\gamma^0\,S(C)\,\bar{\psi}_\beta^\trans(x')
\end{equation}
we get
\begin{equation}\label{4.21}
\CPg \circ \CPg\,:\;
\psi_\alpha(x) \rightarrow
  -\left( U\,U^\ast \right)_{\alpha\beta} \psi_\beta(x)\,.
\end{equation}
We shall, therefore, require
\begin{equation}\label{4.22}
-U\,U^\ast = e^{i\varkappa} \unitmatrix_2
\end{equation}
with real $\varkappa$.
In appendix~\ref{app-mixings} we show that there are only two types of
solutions of \eqref{4.22}.
\begin{align}
\label{4.23}
  \text{Type (a)\,:}\quad
  e^{i\varkappa} &=1\,,
\notag\\
  U\,U^\ast &= -\unitmatrix_2\,,
\notag\\
  U &= e^{i\xi}\epsilon
     = e^{i\xi}\begin{pmatrix} 0 & 1\\ -1 & 0 \end{pmatrix}\,.
\\[2mm]
\label{4.24}
  \text{Type (b)\,:}\quad
  e^{i\varkappa} &=-1\,,
\notag\\
  U\,U^\ast &= \unitmatrix_2\,,
\notag\\
  U &= e^{i\xi}
       \begin{pmatrix} \alpha & \beta \\ \beta & -\alpha^\ast \end{pmatrix}\,,\notag\\
  \beta &\ge 0\,,\quad \abs{\alpha}^2+\beta^2 = 1\,.
\end{align}
This classification of the fermion generation mixings in the \CP\ transformations
in type~(a) and (b) is similar to the Higgs sector, where we distinguish
the \CPg\ transformations of type~(i) and (ii) according to the type of Higgs flavour
mixing involved.
If only the ``standard'' mixing types (ii) respectively (b) occur in a \CPg\ transformation,
the operator $(\CPg)^2$ is normalised as for the standard \CPs\ transformation.
For certain combinations involving the ``non-standard'' mixing types~(i) respectively (a),
the operator $(\CPg)^2  \circ \exp(i 6\pi Y)$ is normalised in the usual way,
with $Y$ being the hypercharge operator;
in other cases additional unobservable minus signs may occur for the fermions.

Let us next note the change of the matrices $U_{R,L}$ of \eqref{4.17}
under a basis change of the fermion fields.
Consider for instance a basis change of the $d_R$ fields as in \eqref{4.15}:
\begin{equation}\label{4.24a}
d_{\alpha R}(x) = V^\dagger_{\alpha\beta}\,d'_{\beta R}(x)\,,
\end{equation}
with $V = (V_{\alpha\beta}) \in U(2)$.
The \CPg\ transformation of $d_{\alpha R}$ following from 
\eqref{4.17}
 is then
\begin{equation}\label{4.24b}
d_{\alpha R}(x) \rightarrow
  \tilde{U}^{(d)}_{R\,\alpha\beta}\,
  \gamma^0\, S(C)\, \bar{d}^\trans_{\beta R}(x')\,,
\end{equation}
with
\begin{equation}\label{4.24c}
\tilde{U}^{(d)}_R = V^\dagger\, U^{(d)}_R\, V^\ast\,.
\end{equation}
It is easy to check that this transformation does not change the type of the
\CPg\ transformation as described by \eqref{4.23} respectively
\eqref{4.24}.

For changes of basis of the other fermion fields the corresponding $U$ matrices
transform analogously to \eqref{4.24c}.

Finally we consider a generic coupling of a fermion doublet field 
$\psi_{L\,\alpha}(x)$ and singlet field $\chi_{R\,\alpha}(x)$ to the Higgs
fields:
\begin{equation}\label{4.25}
\mathscr{L}'(x) = -\bar{\chi}_{\alpha R}(x)\,C^{(i)}_{\alpha\beta}\,
  \varphi_i^\dagger(x)\,\psi_{\beta L}(x)
  + h.c.
\end{equation}
A generic \CPg\ transformation for the Higgs fields as in \eqref{4.16} and
for the fermions according to
\begin{alignat}{2}\label{4.26}
\CPg\,:\quad
&&\psi_{\alpha\,L}(x) \rightarrow
   U^{(\psi)}_{L\,\alpha\beta}\,\gamma^0\,S(C)\,
   \bar{\psi}_{\beta\,L}^\trans(x')\,,
\notag\\
&&\chi_{\alpha\,R}(x) \rightarrow
   U^{(\chi)}_{R\,\alpha\beta}\,\gamma^0\,S(C)\,
   \bar{\chi}_{\beta\,R}^\trans(x')
\end{alignat}
leads to
\begin{align}\label{4.27}
\CPg\,:
\mathscr{L}'(x) \rightarrow &
  -\bar{\chi}_{\alpha\,R}(x')\,\tilde{C}^{(i)}_{\alpha\beta}\,
    \varphi_i^\dagger(x')\,\psi_{\beta\,L}(x')\notag\\
	&+ h.c.
\end{align}
where
\begin{equation}\label{4.28}
\tilde{C}^{(i)} =
   U_R^{(\chi)\,\trans}\,C^{(j)\,\ast}\,U_L^{(\psi)\,\ast}\,W_{ji}\,.
\end{equation}
Similarly we find for a coupling
\begin{equation}\label{4.29}
\mathscr{L}''(x) = \bar{\chi}_{\alpha\,R}(x)\,C'^{(i)}_{\alpha\beta}\,
  \varphi_i^\trans(x)\,\epsilon\,\psi_{\beta\,L}(x)
  + h.c.
\end{equation}
the transformation
\begin{align}\label{4.30}
\CPg\,:
\mathscr{L}''(x) \rightarrow&
  \bar{\chi}_{\alpha\,R}(x')\,\tilde{C}'^{(i)}_{\alpha\beta}\,
    \varphi_i^\trans(x')\,\epsilon\,\psi_{\beta\,L}(x')\notag\\
	&+ h.c.
\end{align}
Here we have
\begin{equation}\label{4.31}
\tilde{C}'^{(i)} = 
   U_R^{(\chi)\,\trans}\,C'^{(j)\,\ast}\,U_L^{(\psi)\,\ast}\,W^\ast_{ji}\,.
\end{equation}

%%%%%%%%%%%%%%%%%%%%%%%%%%%%%%%%%%%%%%%%%%%%%%%%%%%%%%%%%%%%%%%%%%%%%%%%%%%%%%
\subsection{Invariant couplings for two lepton families}
\label{invariantcouplings}

Now we impose our principle of maximal \CP\ invariance on the Yukawa 
interaction \eqref{4.9}.
We want to find out what this implies for the coupling matrices $C^{(j)}_l$, 
$C'^{(j)}_q$ and $C^{(j)}_q$.
We start by considering only the leptonic part of $\mathscr{L}_{\mathrm{Yuk}}$
in \eqref{4.9},
\begin{align}\label{4.32}
\mathscr{L}_{\mathrm{Yuk},l}(x) =&
  -\bar{l}_{\alpha R}(x)\,C^{(j)}_{l\,\alpha\beta}\,\varphi_j^\dagger(x)\,
    \begin{pmatrix} \nu_{\beta L}(x) \\ l_{\beta L}(x) \end{pmatrix}\notag\\
 & + h.c.
\end{align}
As explained in section~\ref{familiesgeneralities} we can, without loss of
generality, suppose \eqref{4.11} to hold.
Now we consider a generic \CPg\ transformation for which \eqref{4.16} holds for 
the Higgs fields.
This \CPg\ can be extended to an invariance of
$\mathscr{L}_{\mathrm{Yuk},l}(x)$ if and only if we can find $U(2)$ matrices
$U^{(l)}_R$ and $U^{(l)}_L$ in \eqref{4.17} 
such that, according to \eqref{4.28}, we have
\begin{equation}\label{4.33}
U^{(l)\,\trans}_R \, C^{(j)\,\ast}_l U^{(l)\,\ast}_L W_{ji} = C^{(i)}_l\,.
\end{equation}
For the principle of maximal \CP\ invariance to hold we must be able to find 
matrices $U^{(l)}_R,~U^{(l)}_L$ solving \eqref{4.33} for all four 
transformations \CPi, \CPa, \CPb\ and \CPc, with the corresponding $W_{ji}$
from Tab.~\ref{tab-cphiggs}.

Let us first consider the case
\begin{equation}\label{4.34}
c^{(1)}_{l\,2} > 0\,,\quad c^{(1)}_{l\,3} > 0\,,\quad
c^{(1)}_{l\,2} \neq c^{(1)}_{l\,3}\,.
\end{equation}
This corresponds to non-vanishing and unequal masses for the leptons $l_2$ and 
$l_3$ after EWSB.
As we show in appendix~\ref{app-leptons} we have, if \eqref{4.34} holds,
solutions of \eqref{4.33} for all four \CPg\ transformations only if the matrix $C^{(2)}_l$ 
has the following structure:
\begin{equation}\label{4.35}
C^{(2)}_l =
  \begin{pmatrix} 0 & C^{(2)}_{l\,23}\\ C^{(2)}_{l\,32} & 0 \end{pmatrix}\,.
\end{equation}
The possible values for $C^{(2)}_{l\,23}$ and $C^{(2)}_{l\,32}$ are listed in
Tab.~\ref{tab-couplppneq} (see \eqref{B.45a}, \eqref{B.45b}, \eqref{B.51} and
\eqref{B.52}).

\begin{table}[htb]
\begin{center}
\begin{tabular}{|c|c|}
\hline
$C^{(2)}_{l23}$&$C^{(2)}_{l32}$\\
\hline\hline
$c^{(1)}_{l3}$&$c^{(1)}_{l2}$\\
$c^{(1)}_{l3}$&$-c^{(1)}_{l2}$\\
$c^{(1)}_{l2}$&$c^{(1)}_{l3}$\\
$c^{(1)}_{l2}$&$-c^{(1)}_{l3}$\\
\hline
\end{tabular}
\caption{\label{tab-couplppneq}
The possible values for $C^{(2)}_{l\,23}$ and $C^{(2)}_{l\,32}$
for the case that \eqref{4.34} holds.
}
\end{center}
\end{table}

\begin{figure}[h!]
\includegraphics[width=0.55\columnwidth]{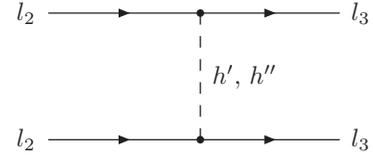}
\caption{\label{lepprocess} Two Feynman diagrams for
the large FCNC process
\mbox{$l_2 + l_2 \rightarrow l_3 + l_3$}
reflecting the last two contributions in 
the Lagrangian~(\ref{4.36}).}
\end{figure}

\begin{figure}[h!]
\includegraphics[width=0.8\columnwidth]{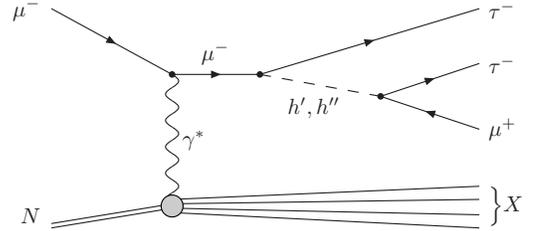}
\caption{\label{disfigure}Two Feynman diagrams for the deep inelastic muon--nucleon
scattering process which would reveal FCNCs 
corresponding to the couplings in~Fig.~\ref{lepprocess}.}
\end{figure}

The corresponding matrices $U^{(l)}_{R,L}$ for all four \CPg\ symmetries are 
given in appendix~\ref{app-leptons}.
To see the physical consequences of this result we look at 
$\mathscr{L}_{\text{Yuk},l}$ after EWSB.
Inserting for the Higgs fields the physical expressions \eqref{3.7a} and 
\eqref{3.8} we get from \eqref{4.32}
\begin{align}\label{4.36}
&\mathscr{L}_{\mathrm{Yuk},l} =\notag\\
&-c^{(1)}_{l\,2}\frac{1}{\sqrt{2}}\left(v_0+\rho'(x)\right)\bar{l}_2(x)\,l_2(x)
\notag\\ &
-c^{(1)}_{l\,3}\frac{1}{\sqrt{2}}\left(v_0+\rho'(x)\right)\bar{l}_3(x)\,l_3(x)
\notag\\ &
-H^-(x) \left[ C^{(2)}_{l\,23}\,\bar{l}_2(x) \omega_L \nu_3(x) \right.\notag\\
&\qquad \qquad \left. +C^{(2)}_{l\,32}\,\bar{l}_3(x) \omega_L \nu_2(x) \right]
\notag\\ &
-H^+(x) \left[ C^{(2)}_{l\,23}\,\bar{\nu}_3(x) \omega_R l_2(x) \right.\notag\\
&\qquad \qquad \left.  +C^{(2)}_{l\,32}\,\bar{\nu}_2(x) \omega_R l_3(x) \right]
\notag\\ &
  -\frac{h'(x)}{\sqrt{2}} \left\{
    \bar{l}_2(x)
    \left[ C^{(2)}_{l\,23} \omega_L + C^{(2)}_{l\,32} \omega_R \right]
    l_3(x)
  \right.
\notag\\ &
  \left. \phantom{-\frac{h'(x)}{\sqrt{2}}}
    +\bar{l}_3(x)
    \left[ C^{(2)}_{l\,23} \omega_R + C^{(2)}_{l\,32} \omega_L \right]
    l_2(x)
  \right\}
\notag\\ &
  -\frac{i\,h''(x)}{\sqrt{2}}
  \left\{
    \bar{l}_2(x)
    \left[ -C^{(2)}_{l\,23} \omega_L + C^{(2)}_{l\,32} \omega_R \right]
    l_3(x)
  \right.
\notag\\ &
  \left. \phantom{-\frac{i\,h''(x)}{\sqrt{2}}}
    +\bar{l}_3(x)
    \left[ C^{(2)}_{l\,23} \omega_R - C^{(2)}_{l\,32} \omega_L \right]
    l_2(x)
  \right\}
\end{align}
with the chirality projectors 
\begin{equation}
\label{projectors}
\omega_R:=\frac{1+\gamma_5}{2}
\quad\text{and}\quad
\omega_L:=\frac{1-\gamma_5}{2}.
\end{equation}
Here we can read off the lepton masses
\begin{align}\label{4.37}
m_{l\,2} &= c^{(1)}_{l\,2}\,\frac{v_0}{\sqrt{2}}\,,
\notag\\
m_{l\,3} &= c^{(1)}_{l\,3}\,\frac{v_0}{\sqrt{2}}\,.
\end{align}
Identifying the lepton 3 with the $\tau$ lepton we see that in all cases listed 
in Tab.~\ref{tab-couplppneq} either $\abs{C^{(2)}_{l\,23}}=m_\tau\sqrt{2}/v_0$
or $\abs{C^{(2)}_{l\,32}}=m_\tau\sqrt{2}/v_0$.
Thus \eqref{4.36} always contains large lepton flavour-changing neutral 
currents, FCNCs.
These would allow for processes like
\begin{equation}\label{4.38}
l_2 + l_2 \rightarrow l_3 + l_3
\end{equation}
through diagrams like in~Fig.~\ref{lepprocess}.
A direct study of process \eqref{4.38}, $\mu^- + \mu^- \to \tau^- + \tau^-$,
would be a topic for a muon collider which, however, is far in the future.
But the couplings in~Fig.~\ref{lepprocess} would also lead to
spectacular lepton-flavour-violating events in deep inelastic muon--nucleon
scattering,
\begin{equation}\label{4.38a}
\mu^- + N \rightarrow \mu^+ + \tau^- + \tau^- + X\,.
\end{equation}
Two of the corresponding tree level Feynman diagrams are shown in~Fig.~\ref{disfigure}.
Here $X$ stands for the hadronic final state.
Since such FCNCs were never observed
we consider them to be unacceptable phenomenologically.

The next case to study is
\begin{equation}\label{4.39}
c^{(1)}_{l\,2} = c^{(1)}_{l\,3} > 0\,.
\end{equation}
There we can construct a coupling \eqref{4.32} satisfying the principle of 
maximal \CP\ invariance and having no \mbox{FCNCs}.
We give the details in appendix~\ref{app-leptons}.
However, here we have, according to \eqref{4.37} equal lepton masses,
$m_{l\,2}=m_{l\,3}$, which is, again, unacceptable phenomenologically.

It remains to be seen what happens for the case of one massless and one 
massive lepton.
Taking, by convention, $l_3$ to be the massive lepton we have the case
\begin{equation}\label{4.40}
c^{(1)}_{l\,2} = 0\,,\qquad c^{(1)}_{l\,3} > 0\,.
\end{equation}
Here we shall {\em prescribe} the form of the matrices $U^{(l)}_R$ and 
$U^{(l)}_L$ for the four \CPg\ transformations as shown in 
Tab.~\ref{tab-cpzp}.
\CPb\ is the standard \CP\ transformation for all fields, $\CPb = \CPs$.
We require now invariance of the Yukawa interaction \eqref{4.32} under these 
four \CPg\ transformations, that is, we require \eqref{4.33} to hold
with the corresponding $W_{ij}$ from Tab.~\ref{tab-cpzp}.
It is easy to check that starting with $c^{(1)}_{l\,2} \geq 0$,
$c^{(1)}_{l\,3} > 0$ instead of \eqref{4.40} these invariances {\em require}
$c^{(1)}_{l2}=0$ and 
\begin{equation}\label{4.41}
C^{(2)}_l = \begin{pmatrix} -c^{(1)}_{l\,3} & 0 \\ 0 & 0 \end{pmatrix}\,;
\end{equation}
see \eqref{C.45}.
The resulting Yukawa term reads
\begin{align}\label{4.42}
\mathscr{L}_{\mathrm{Yuk},l}(x) =
 - c^{(1)}_{l\,3} &\left\{ 
   \bar{l}_{3\,R}(x)\,\varphi_1^\dagger(x)
   \begin{pmatrix} \nu_{3\,L}(x) \\ l_{3\,L}(x) \end{pmatrix}
   \right.
\notag\\
  &\left.
  -\bar{l}_{2\,R}(x)\,\varphi_2^\dagger(x)
   \begin{pmatrix} \nu_{2\,L}(x) \\ l_{2\,L}(x) \end{pmatrix}
		\right\}
	+ h.c.
\end{align}
Note the high degree of symmetry between the families here.
However, after EWSB we get, inserting \eqref{3.7a} and \eqref{3.8} for the 
Higgs fields and using \eqref{4.37},
\begin{align}\label{4.43}
\mathscr{L}_{\mathrm{Yuk},l}(x) &=
  -m_{l\,3}\left( 1 + \frac{\rho'(x)}{v_0} \right) \bar{l}_3(x)\,l_3(x)
\notag\\
&\quad
  +\frac{m_{l\,3}}{v_0}\, h'(x)\,\bar{l}_2(x)\,l_2(x)
\notag\\
&\quad
  +i \frac{m_{l\,3}}{v_0}\, h''(x)\,\bar{l}_2(x) \gamma_5 l_2(x)
\notag\\
&\quad
  +\frac{\sqrt{2}\,m_{l\,3}}{v_0}\left[
    H^+(x)\, \bar{\nu}_2(x) \omega_R l_2(x)\right.\notag\\
&\quad\qquad\qquad	\left. +H^-(x)\, \bar{l}_2(x) \omega_L \nu_2(x) \right]\,.
\end{align}
The lepton $l_3$ has become massive and couples to the physical $\rho'$ Higgs.
The lepton $l_2$ is massless but couples to $h',h''$ and the charged Higgs bosons
$H^\pm$.
\begin{table}
\begin{center}
\begin{tabular}{|l|c|c|c|}
\hline
\CPg & W & $U^{(l)}_R$&$U^{(l)}_L$\\
\hline\hline
\CPi & $\epsilon$   & $\epsilon$   & $\sigma^1$\\
\CPa & $\sigma^3$      & $-\sigma^3$     & $\unitmatrix_2$\\
\CPb & $\unitmatrix_2$ & $\unitmatrix_2$ & $\unitmatrix_2$\\
\CPc & $\sigma^1$      & $-\sigma^1$     & $\sigma^1$\\
\hline
\end{tabular}
\caption{\label{tab-cpzp}
The matrices $W$, $U^{(l)}_R$ and $U^{(l)}_L$ for the four \CPg\ transformations
for the case of one massless and one massive lepton; see~\eqref{4.40}.
}
\end{center}
\end{table}

In appendix~\ref{app-leptons} we give a general discussion of the case where
\eqref{4.40} holds; that is, where lepton $l_2$ is massless and lepton $l_3$
massive.
We show there that the requirements of maximal \CP\ invariance plus absence of
FCNCs uniquely leads to the coupling \eqref{4.42}.

%%%%%%%%%%%%%%%%%%%%%%%%%%%%%%%%%%%%%%%%%%%%%%%%%%%%%%%%%%%%%%%%%%%%%%%%%%%%%%
\subsection{Invariant couplings for two quark families}
\label{quarkfamilies}

In this section we study the quark part of the Lagrangian \eqref{4.9}.
Let us first look at the term which generates masses for the $u$-type quarks,
\begin{align}\label{4.44}
\mathscr{L}_{\mathrm{Yuk},q'}(x)=&
  \bar{u}_{\alpha\,R}(x) C'^{(j)}_{q\,\alpha\beta}\, \varphi_j^\trans(x)\,
    \epsilon \begin{pmatrix} u_{\beta\,L}(x)\\ d'_{\beta\,L}(x) \end{pmatrix}\notag\\
	&+ h.c.
\end{align}
Here we can suppose without loss of generality that $C'^{(1)}_q$ is as in 
\eqref{4.12}.
As for the case of the leptons in section \ref{invariantcouplings} we ask if 
$\mathscr{L}_{\mathrm{Yuk},q'}$ in \eqref{4.44} allows for the implementation of our
principle of maximal \CP\ invariance.
That is, we ask for matrices 
$U^{(u)}_R,~U^{(u)}_L$ in 
\eqref{4.17} 
which satisfy either \eqref{4.23} or 
\eqref{4.24} and allow us to solve (see \eqref{4.31})
\begin{equation}\label{4.45}
U^{(u)\,\trans}_R\, C'^{(j)\,\ast}_q\, U^{(u)\,\ast}_L\, W^\ast_{ji} =
  C'^{(i)}_q\,,
\end{equation}
for all four \CP\ symmetries with $W_{ji}$ as given in Tab.~\ref{tab-cphiggs}.
Since all $W$ matrices are real \eqref{4.45} is completely analogous to 
\eqref{4.33}.
We can immediately conclude from the results of section \ref{invariantcouplings} 
that for the case
\begin{equation}\label{4.46}
c^{(1)}_{u\,2} > 0\,,\quad c^{(1)}_{u\,3} > 0\,,\quad
c^{(1)}_{u\,2} \neq c^{(1)}_{u\,3}
\end{equation}
the principle of maximal \CP\ invariance leads to large FCNCs.
Here it is important to note that these FCNCs are generated for the physical 
mass eigenfields $u_2$ and $u_3$.
Requiring the absence of these FCNCs then allows for only two possibilities for a 
non-zero coupling $\mathscr{L}_{\mathrm{Yuk},q'}$.
Either we must have non-zero equal masses for the quarks $u_2$ and $u_3$ or we
must have $u_2$ massless and $u_3$ massive.
Discarding the former for phenomenological reasons we are left with the case of 
a massless $u_2=c$ quark and a massive $u_3=t$ quark.
Now we {\em prescribe} the matrices $U^{(u)}_R$ and $U^{(u)}_L$ for the four 
\CPg\ transformations as for the lepton case in Tab.~\ref{tab-cpzp}.
That is, we set for all \CPg\ transformations
\begin{align}\label{4.47}
U^{(u)}_R &= U^{(l)}_R\,,
\notag\\
U^{(u)}_L &= U^{(l)}_L\,.
\end{align}
With~(\ref{4.47}) we find that the matrices $C'^{(j)}_q~(j=1,2)$ have to be as 
follows:
\begin{align}\label{4.48}
C'^{(1)}_q &= \begin{pmatrix} 0 & 0 \\ 0 & c^{(1)}_{u\,3} \end{pmatrix}\,,
\quad c^{(1)}_{u\,3}>0\,,\notag\\
C'^{(2)}_q &= \begin{pmatrix} -c^{(1)}_{u\,3} & 0 \\ 0 & 0 \end{pmatrix}\,.
\end{align}
The resulting coupling term \eqref{4.44} reads
\begin{align}\label{4.48a}
\mathscr{L}_{\mathrm{Yuk},q'}(x) =
 c^{(1)}_{u\,3} &\left\{
   \bar{u}_{3\,R}(x)\,\varphi_1^\trans(x)\,\epsilon
   \begin{pmatrix} u_{3\,L}(x) \\ d'_{3\,L}(x) \end{pmatrix}
   \right.
\notag\\
  &\left.
  -\bar{u}_{2\,R}(x)\,\varphi_2^\trans(x)\,\epsilon
   \begin{pmatrix} u_{2\,L}(x) \\ d'_{2\,L}(x) \end{pmatrix}
		\right\}
	+ h.c.
\end{align}
As for the case of the leptons (see appendix~\ref{app-leptons}) we can show
the following.
For $c^{(1)}_{u\,2} = 0$, $c^{(1)}_{u\,3} > 0$ the principle of maximal
\CP\ invariance together with the requirement of absence of FCNCs leads
uniquely to the coupling~\eqref{4.48a}.

We turn next to the term in \eqref{4.9} which generates masses for $d$-type 
quarks 
\begin{align}\label{4.49}
\mathscr{L}_{\mathrm{Yuk},q}(x) =&
  -\bar{d}'_{\alpha\,R}(x)\,C^{(j)}_{q\,\alpha\beta}\, 
    \varphi_j^\dagger(x)\,
   \begin{pmatrix} u_{\beta\,L}(x) \\ d'_{\beta\,L}(x) \end{pmatrix}\notag\\
	&  + h.c.
\end{align}
Here the standard form for $C^{(1)}_q$ is given in \eqref{4.13} and 
\eqref{4.14}.
Note that $d'_\beta$ are -- in general -- not the mass eigenfields.
We shall change to the basis of $d_\alpha$ mass eigenfields and the 
corresponding isospin partners of $u'_\alpha$ fields according to \eqref{4.15}
setting
\begin{align}\label{4.50}
d_{\alpha\,R}(x) &= V^\dagger_{\alpha\beta}\,d'_{\beta\,R}\,,
\notag\\
\begin{pmatrix} u'_{\alpha\,L}(x) \\ d_{\alpha\,L}(x) \end{pmatrix} &=
  V^\dagger_{\alpha\beta} 
   \begin{pmatrix} u_{\beta\,L}(x) \\ d'_{\beta\,L}(x) \end{pmatrix}\,.
\end{align}
The coupling term \eqref{4.49} reads now
\begin{align}\label{4.51}
\mathscr{L}_{\mathrm{Yuk},q}(x)=&
  -\bar{d}_{\alpha\,R}(x)\, \tilde{C}^{(j)}_{q\,\alpha\beta}\,
   \varphi_j^\dagger(x)\,
   \begin{pmatrix} u'_{\beta\,L}(x)\\ d_{\beta\,L}(x) \end{pmatrix}\notag\\
	& + h.c.
\end{align}
where 
\begin{equation}\label{4.52}
\tilde{C}^{(j)}_q = V^\dagger\, C^{(j)}_q\, V,\quad j=1,2\,.
\end{equation}
From \eqref{4.13} we see that this implies 
\begin{equation}\label{4.53}
\tilde{C}^{(1)}_q =
  \begin{pmatrix} c^{(1)}_{d\,2} & 0 \\ 0 & c^{(1)}_{d\,3} \end{pmatrix}\,.
\end{equation}

Now we can proceed as for the lepton case.
We see that requiring the principle of maximal \CP\ invariance leads for the
case
\begin{equation}\label{4.54}
c^{(1)}_{d\,2} > 0\,,\quad c^{(1)}_{d\,3} > 0\,,\quad
c^{(1)}_{d\,2} \neq c^{(1)}_{d\,3}
\end{equation}
to large FCNCs among the physical $d$-quark mass eigenfields.
These FCNCs can only be avoided if we require either equal masses 
$m_{d2}=m_{d3}$ or $m_{d2}=0$ and $m_{d3}\neq 0$.
Again we discard the former possibility for phenomenological reasons.
For the case $m_{d2}=0,~m_{d3}\neq 0$ we shall again {\em prescribe} the \CPg\ 
transformation matrices of the fermion fields.
But we have to remember that we have already prescribed the transformation 
matrices $U^{(u)}_L$ for the left-handed quark doublets in \eqref{4.47}.
Since the same doublets appear in \eqref{4.49} we can here only prescribe 
$U^{(d)}_R$ in 
%\eqref{4.18}, 
\eqref{4.17}, 
since everything else is already fixed.
Note that $U^{(d)}_R$ refers again to the $d'_\alpha$ fields.
We prescribe here 
\begin{equation}\label{4.55}
U^{(d)}_R = U^{(l)}_R
\end{equation}
as in Tab.~\ref{tab-cpzp}.
This leads to 
\begin{equation}\label{4.56}
V = \unitmatrix_2\,,
\end{equation}
that is, to a CKM matrix equal to unity in the 2--3 sector as we show in 
appendix~\ref{app-quarks}.
The final form of the coupling term $\mathscr{L}_{\mathrm{Yuk},q}$ is as follows:
\begin{align}\label{4.57}
\mathscr{L}_{\mathrm{Yuk},q}(x) =
 -c^{(1)}_{d\,3} &\;\Bigg\{
   \bar{d}_{3\,R}(x)\,\varphi_1^\dagger(x)
   \begin{pmatrix} u_{3\,L}(x) \\ d_{3\,L}(x) \end{pmatrix}
\notag\\
  &-\bar{d}_{2\,R}(x)\,\varphi_2^\dagger(x)
   \begin{pmatrix} u_{2\,L}(x) \\ d_{2\,L}(x) \end{pmatrix}
		\Bigg\}
	+ h.c.
\notag\\[-5pt]
\end{align}
Note that with \eqref{4.56} we have $d'_{\alpha R,L}=d_{\alpha R,L}$ for the 
$d$-type fields.

In appendix~\ref{app-quarks} we give a general discussion of the case
$c^{(1)}_{d\,2} = 0$, $c^{(1)}_{d\,3} > 0$, that is of the case where $d_2$ is
massless and $d_3$ massive.

%%%%%%%%%%%%%%%%%%%%%%%%%%%%%%%%%%%%%%%%%%%%%%%%%%%%%%%%%%%%%%%%%%%%%%%%%%%%%%
\section{Discussion}
%%%%%%%%%%%%%%%%%%%%%%%%%%%%%%%%%%%%%%%%%%%%%%%%%%%%%%%%%%%%%%%%%%%%%%%%%%%%%%
\label{sec-discussion}

In this section we collect the results found in the previous sections
and subsequently discuss their physical consequences.
We have investigated a two-Higgs-doublet model having four 
generalised \CP\ transformations as symmetries.
We have introduced the principle of maximal \CP\ invariance
which requires that these four symmetries are extendable to the full Lagrangian.

In sections~\ref{sec-model} and \ref{sec-higgses} we have discussed the Higgs 
sector of the model which is characterised by the requirement of \CPi\ 
invariance.
We have seen that this leads automatically to three more \CPg\ invariances, 
\CPa, \CPb\ and \CPc.
The EWSB breaks \CPi\ and \CPc\ spontaneously.
At tree level, which we have discussed in this paper, the symmetries 
\CPa\ and \CPb\ are unbroken.

In section~\ref{sec-fermions} we studied if we can implement the principle of 
maximal \CP\ invariance, that is, if our four \CPg\ symmetries can be extended
to the coupling of the fermions to the Higgs fields.
For this we introduced the fermion families; see Tab.~\ref{tab-fermions},
taking the neutrinos as massless.
We found in section~\ref{sec-onefamily} that requiring a single family to have
a \CPi\ invariant coupling leads necessarily to the coupling being identically
zero.
Thus, we have the interesting conclusion that a single fermion family with 
non-zero couplings and, therefore, masses is not consistent with \CPi\ 
invariance, and therefore, a forteriori, with the principle of maximal \CP\ 
invariance.

In sections \ref{familiesgeneralities} to \ref{quarkfamilies} we discussed 
non-zero couplings of two families to the Higgs doublets, always requiring the 
principle of maximal \CP\ invariance.
We took the two families to be the second and the third.
We found that unequal non-zero masses for the leptons $l_2$ and $l_3$,
the quarks $u_2$ and $u_3$, as well as $d_2$ and $d_3$ always implied large 
flavour-changing neutral currents (FCNCs).
The absence of large FCNCs required either equal masses of corresponding 
fermions $(m_{l2}=m_{l3}$ etc.) or one fermion massless, the other massive.
Discarding the equal mass case on phenomenological grounds we were, thus, left 
with the possibility 
\begin{align}\label{5.1}
m_{l\,2} &= 0\,, &  m_{l\,3} &\neq 0\,,
\notag\\
m_{u\,2} &= 0\,, &  m_{u\,3} &\neq 0\,,
\notag\\
m_{d\,2} &= 0\,, &  m_{d\,3} &\neq 0\,.
\end{align}
The specific set of \CP\ symmetries defined by
%\eqref{4.18} 
\eqref{4.17} 
together with Tab.~\ref{tab-cpzp}, (\ref{4.47}) and (\ref{4.55}) 
guarantees the absence of large FCNCs and requires
vanishing masses for family 2.
As a further consequence the CKM matrix between the families 2 and 3
has to be equal to unity; see \eqref{4.56}.
Combining \eqref{4.42}, \eqref{4.48a} and \eqref{4.57} we find for the
full Yukawa part of the Lagrangian the simple form
\begin{align}\label{5.2}
\mathscr{L}_{\mathrm{Yuk}}(x) = 
  -c^{(1)}_{l\,3} & \;\Bigg\{
    \bar{l}_{3\,R}(x)\,\varphi_1^\dagger(x)
    \begin{pmatrix} \nu_{3\,L}(x) \\ l_{3\,L}(x) \end{pmatrix}
\notag\\ &
    -\bar{l}_{2\,R}(x)\,\varphi_2^\dagger(x)
    \begin{pmatrix} \nu_{2\,L}(x) \\ l_{2\,L}(x) \end{pmatrix}
    \Bigg\}
\notag\\
  +c^{(1)}_{u\,3} &\;\Bigg\{
    \bar{u}_{3\,R}(x)\,\varphi_1^\trans(x)\,\epsilon
    \begin{pmatrix} u_{3\,L}(x) \\ d_{3\,L}(x) \end{pmatrix}
\notag\\ &
    -\bar{u}_{2\,R}(x)\,\varphi_2^\trans(x)\,\epsilon
    \begin{pmatrix} u_{2\,L}(x) \\ d_{2\,L}(x) \end{pmatrix}
    \Bigg\}
\notag\\
  -c^{(1)}_{d\,3} &\;\Bigg\{
    \bar{d}_{3\,R}(x)\,\varphi_1^\dagger(x)
    \begin{pmatrix} u_{3\,L}(x) \\ d_{3\,L}(x) \end{pmatrix}
\notag\\ &
    -\bar{d}_{2\,R}(x)\,\varphi_2^\dagger(x)
    \begin{pmatrix} u_{2\,L}(x) \\ d_{2\,L}(x) \end{pmatrix}
    \Bigg\}
		+ h.c.
\end{align}
In this model the first family remains uncoupled to the Higgs fields.

For the convenience of the reader we summarise here the generalised \CP\
symmetries of the full Lagrangian~(\ref{2.2}) with three generations of fermions,
the Higgs part given by~(\ref{2.3}) and (\ref{2.20}), and the Yukawa
term given by~(\ref{5.2}). 
For any \CPg\ we transform the gauge bosons as in~(\ref{2.13}) and the
first generation fermions as in~(\ref{4.2}) where $\xi_1$ to $\xi_5$
remain arbitrary. The Higgs fields are transformed according to~(\ref{4.16})
and the second and third generation fermions according to~(\ref{4.17}).
The matrices $W$ in~(\ref{4.16}) and $U_L^{(l)}$ to
$U_R^{(d)}$ in~(\ref{4.17}) are summarised in Tab.~\ref{tab-5}.
In appendix~\ref{app-D} we discuss the relation of these
\CPg\ invariances to conventional discrete symmetries.

After EWSB we insert the Higgs fields parametrised by the physical fields as 
in \eqref{3.7a} and \eqref{3.8} and use the relations
\begin{align}\label{5.3}
m_{l\,3} &= c^{(1)}_{l\,3}\frac{v_0}{\sqrt{2}} \equiv m_\tau\,,
\notag\\
m_{u\,3} &= c^{(1)}_{u\,3}\frac{v_0}{\sqrt{2}} \equiv m_t\,,
\notag\\
m_{d\,3} &= c^{(1)}_{d\,3}\frac{v_0}{\sqrt{2}} \equiv m_b\,.
\end{align}
We find then from~(\ref{5.2}) with $\omega_{R,L}$ defined in~(\ref{projectors})
\begin{multline}\label{5.4}
\mathscr{L}_{\mathrm{Yuk}}(x) =\\
\shoveleft{\begin{alignedat}{1}
&\quad
  -m_{l\,3} \bigg( 1 + \frac{\rho'(x)}{v_0} \bigg) \bar{l}_3(x)\,l_3(x)
\\ &\quad
  -m_{u\,3} \bigg( 1 + \frac{\rho'(x)}{v_0} \bigg) \bar{u}_3(x)\,u_3(x)
\\ &\quad
  -m_{d\,3} \bigg( 1 + \frac{\rho'(x)}{v_0} \bigg) \bar{d}_3(x)\,d_3(x)
\\
\end{alignedat}}
\\
\shoveleft{\begin{alignedat}{2}
&\quad
  + \frac{h'(x)}{v_0} &&\bigg[\;\;\,
    m_{l\,3}\, \bar{l}_2(x)\,l_2(x)\\
   &&&+m_{u\,3}\, \bar{u}_2(x)\,u_2(x)\\
   &&&+m_{d\,3}\, \bar{d}_2(x)\,d_2(x) \bigg]
\\ &\quad
	+ i \, \frac{h''(x)}{v_0} &&\bigg[\;\;\,
    m_{l\,3}\, \bar{l}_2(x)\, \gamma_5\, l_2(x)\\
   &&&-m_{u\,3}\, \bar{u}_2(x)\, \gamma_5\, u_2(x)\\
   &&&+m_{d\,3}\, \bar{d}_2(x)\, \gamma_5\, d_2(x) \bigg]
\\
\end{alignedat}}
\\
\shoveleft{\begin{alignedat}{2}
&\quad
+\bigg\{ \frac{H^+(x) \sqrt{2}}{v_0}&&\bigg[\;\;\,
     m_{l\,3}\,\bar{\nu}_2(x)\, \omega_R\, l_2(x)\\
   &&& -m_{u\,3}\,\bar{u}_2(x)\,\omega_L\,d_2(x)\\
   &&& +m_{d\,3}\,\bar{u}_2(x)\,\omega_R\,d_2(x)
  \bigg] + h.c. \bigg\}\,.
\end{alignedat}}
\end{multline}

We discuss now the transformation properties of the
physical fields after EWSB under the \CPg\ transformations
of~Tab.~\ref{tab-5}.
The symmetries \CPi\ and \CPc\ are spontaneously broken.
Thus they are not explicitly visible for the physical
fields.
The symmetries \CPa\ and \CPb\ are unbroken and are thus
directly reflected by the physical fields.
The transformation of the gauge bosons is always given
by~(\ref{2.13}); see the remark after~(\ref{2.16a}).
The transformations of the physical Higgs fields under
\CPa\ and \CPb\ are given in~(\ref{3.23}) and (\ref{3.24}), respectively.
For the first generation fermions both, \CPa\ and \CPb,
can be taken to be the standard \CP\ transformation,
setting $\xi_1=\ldots=\xi_5=0$ in~(\ref{4.2}).
The transformation \CPb\ acts as standard \CP\ transformation
also for the second and third fermion families; see~Tab.~\ref{tab-5}.
The transformation \CPa\ acts, according to~Tab.~\ref{tab-5},
on the second and third generation lepton fields
as follows; see (\ref{4.17}):
\begin{alignat}{2} \label{122a}
\CPa\,:\;
&& \nu_{\mu L}(x) & \rightarrow \phantom{+}\gamma^0 S(C)\bar{\nu}^\trans_{\mu L} (x')\;,  \notag\\
&& \mu_{L}(x) & \rightarrow \phantom{+}\gamma^0 S(C) \bar{\mu}^\trans_{L} (x')\;,  \notag\\
&& \mu_{R}(x) & \rightarrow - \gamma^0 S(C) \bar{\mu}^\trans_{R} (x')\;,  \notag\\
&& \nu_{\tau L}(x) & \rightarrow \phantom{+}\gamma^0 S(C)\bar{\nu}^\trans_{\tau L} (x')\;,  \notag\\
&& \tau(x) & \rightarrow \phantom{+}\gamma^0 S(C)\bar{\tau}^\trans(x')\;.
\end{alignat}
Thus the \CPa\ transformations for the states of $\nu_\mu$, $\nu_\tau$ and $\tau$
are as for the standard \CP\ transformation. For the muons, however, we
have at the level of the states
\begin{alignat}{2} \label{122b}
\CPa\,:\;&&
 \left|\mu^- (\tvec{k},s) \right\rangle  & \rightarrow (-1)^{s+1/2}\;
 \left|\mu^+(-\tvec{k},-s)\right\rangle\;,\notag\\
\CPb\,:\;&&
 \left|\mu^- (\tvec{k},s) \right\rangle  & \rightarrow \left|\mu^+(-\tvec{k},-s)\right\rangle\;.
\end{alignat}
Here $\tvec{k}$ is the momentum and $s=\pm1/2$ is the helicity of the
state. Thus \CPa\ differs from the standard \CP\ transformation \CPb\ by an extra minus sign in the transformation of the right-handed $\mu^- (s=1/2)$. Note that in our
theory as it has been developed so far the muon is massless. Thus its helicity is a
Lorentz-invariant quantity. For the second and third generation quarks the
transformations \CPa\ and \CPb\ act analogously to the lepton case.

Finally we stress again that the theory defined  -- before EWSB --
by the Lagrangian~(\ref{2.2}) with $\mathscr{L}_\varphi$
given by~(\ref{2.3}) and (\ref{2.20})-(\ref{2.22}) and
$\mathscr{L}_\text{Yuk}$ given by (\ref{5.2}) is symmetric under all
four \CPg\ transformations as defined in~Tab.~\ref{tab-5}. Moreover, it
is the most general theory with these symmetries. That is, there is no
further symmetric renormalisable term which could be added. 

%Now we come to the physical discussion of our model.
We consider it noteworthy that our symmetry principles require more than one 
family.
For two families we get in a natural way mass hierarchies.
Choosing the simplest extension to three families we get masses unequal to zero 
only for $\tau,t$ and $b$ whereas all other leptons and quarks, $\mu,e,c,u,s,d$ 
stay massless.
In addition, the CKM matrix of the quarks equals the unit matrix,
$V={\mathbbm 1}$.
Clearly, all this is not quite as one observes it in Nature.
On the other hand, as a first approximation, it is also not so bad.
We have~\cite{Yao:2006px, Jamin:2006tj, Brambilla:2004wf}
\begin{alignat}{4}\label{5.5}
&\frac{m_e}{m_\tau} &&\approx  0.00029\,,
&\qquad&\frac{m_\mu}{m_\tau} &&\approx 0.059\,,
\notag\\
&\left.\frac{m_u}{m_t}\right|_{v_0} &&\approx 9.9\cdot 10^{-6}\,,
&&\left.\frac{m_c}{m_t}\right|_{v_0} &&\approx 0.0036\,,
\notag\\
&\left.\frac{m_d}{m_b}\right|_{v_0} &&\approx 0.0010\,,
&&\left.\frac{m_s}{m_b}\right|_{v_0} &&\approx 0.018\,.
\end{alignat}
Here we have used for the quarks the $\overline{MS}$ masses at the 
renormalisation point \mbox{$\mu=v_0\approx246$~GeV} and \mbox{$\alpha_s(m_Z)=0.119$}.
This electroweak scale seems to us a natural choice for our purpose.
Also the CKM matrix is in Nature not too far from unity.
Indeed, one finds for the absolute values $|V_{ij}|$ \cite{Yao:2006px}
\begin{equation}\label{5.6}
\begin{pmatrix} \abs{V_{11}} & \abs{V_{12}} & \abs{V_{13}} \\
                \abs{V_{21}} & \abs{V_{22}} & \abs{V_{23}} \\
                \abs{V_{31}} & \abs{V_{32}} & \abs{V_{33}} \end{pmatrix} \approx
\begin{pmatrix} 0.974& 0.227& 0.004\\
                0.227& 0.973& 0.042\\
                0.008& 0.042& 0.999\end{pmatrix}.
\end{equation}
Note that in the 2--3 sector $V$ is very close to the unit matrix.
But clearly a good theory should be able to explain the experimental numbers 
in \eqref{5.5} and \eqref{5.6} in more detail.

%\medskip
\begin{table}[tb]
\centering
\begin{tabular}{|c|c|c|c|}
\hline
\;\;\; \CPg \;\;\;& $\;\;W\;\;$ & $U_R^{(l)}=U_R^{(u)}=U_R^{(d)}$ & $U_L^{(l)}=U_L^{(u)}$ \\
\hline\hline
\CPi &$\epsilon$ & $\phantom{+}\epsilon$ & $\sigma^1$\\
\CPa &$\sigma^3$ & $-\sigma^3$ & $\unitmatrix_2$\\
\CPb &$\unitmatrix_2$ & $\phantom{+}\unitmatrix_2$ & $\unitmatrix_2$\\
\CPc &$\sigma^1$ & $-\sigma^1$ & $\sigma^1$\\
\hline
\end{tabular}
\caption{\label{tab-5}
The matrices $W$ \eqref{4.16} and  
$U_L^{(l)}$ to
$U_R^{(d)}$ in~(\ref{4.17}) giving
the four \CPg\ invariances of the Lagrangian
with the Yukawa term~(\ref{5.2}).}
\end{table}
%

%%%%%%%%%%%%%%%%%%%%%%%%%%%%%%%%%%%%%%%%%%%%%%%%%%%%%%%%%%%%%%%%%%%%%%%%%%%%%%
\section{Conclusions}
%%%%%%%%%%%%%%%%%%%%%%%%%%%%%%%%%%%%%%%%%%%%%%%%%%%%%%%%%%%%%%%%%%%%%%%%%%%%%%
\label{sec-conclusions}

We have studied a two-Higgs-doublet model where the scalar sector has four 
generalised \CP\ symmetries. Two of these symmetries 
are spontaneously broken by the 
electroweak symmetry breaking (EWSB).
We have introduced the {\em principle of maximal \CP\ invariance}
requiring that these four \CP\ symmetries can be extended to the
full Lagrangian of the theory. We find that for a single fermion family
this principle forbids a non-zero fermion--Higgs coupling. Thus, if we
want massive fermions which arise from non-zero Yukawa couplings we need
family replication. We have studied then in detail theories with 
two fermion families. Here, indeed, we can extend all four \CP\
symmetries to the full Lagrangian with non-zero Yukawa couplings which are,
however, highly constrained.
Discarding extensions which enforce large flavour-changing neutral currents,
we are left with the possibilities of either equal masses for the
corresponding fermions in the families or large mass hierarchies.
Choosing the latter possibility we arrive at a theory
with a high degree of symmetry between the two families and absence of 
flavour-changing neutral currents.
The Yukawa part of this theory is given in~(\ref{5.2})
and after EWSB in~(\ref{5.4}).
Through EWSB one family becomes massive the other stays massless at the
tree level, which we have discussed in this paper.
We have shown that we can also obtain this theory directly
from a symmetry requirement. For this we prescribe the form
of the four \CP\ transformations for the lepton and quark fields
as shown in~Tab.~\ref{tab-5}.
Our principle of maximal \CP\ invariance leads then directly to the
Yukawa coupling~(\ref{5.2}) implying one massive and one massless family
as well as absence of large FCNCs.
Adding a fermion family uncoupled to the Higgs particles we arrive
at a model which looks like giving a rough approximation of the
structure of fermions observed in Nature. We have massless
neutrinos. Concerning the charged fermions we have one massive family
which we identify with the {\em third} one~($\tau$, $t$, $b$) and two
massless ones which we identify with the {\em second}~($\mu$, $c$, $s$)
and the {\em first}~($e$, $u$, $d$) families. In our model
the CKM matrix between the quark generations is equal to unity.
As for any THDM, the spectrum of physical Higgs particles consists of three
neutral scalars, $\rho'$, $h'$ and $h''$, and the charged Higgs bosons $H^\pm$.
The neutral Higgs particle $\rho'$ -- which has essentially the same properties
as the SM Higgs -- couples exclusively to the third family of fermions.
The other Higgs bosons $h',h''$ and $H^\pm$ couple exclusively to the second family
of fermions. The first fermion family remains uncoupled to the Higgs bosons.
Of course, in reality these statements are expected to be only approximately 
true.
Thus, many open problems remain: Suppose that we start from our
highly \CP\-symmetric theory. How can we obtain masses also for the
first and second fermion families and the \CP\ violation in the CKM matrix?
Can we get the right amount of \CP\ violation to meet the Sakharov
criteria for dynamical generation of the baryon--antibaryon asymmetry in the Universe?
What are the effects of radiative corrections in the theory?
Can the theory be obtained for instance in some grand unified scenario by integrating out
heavy modes?
These questions clearly go beyond the scope of the present paper and must be left
for further studies.
To summarise: we have discussed a two-Higgs-doublet model in which the requirement
of maximal \CP\ invariance provides a mechanism to obtain interesting
structures for fermion masses and couplings.
It remains to be seen if Nature makes use of such a mechanism or if our theory
is only a caricature of reality.
The experiments at the LHC may be able to tell.

\acknowledgments
The authors would like to thank \mbox{R.~Barbieri} and \mbox{C.~Ewerz}
for useful discussions and \mbox{M.~Jamin} for correspondence concerning
the quark mass values.

\appendix
\numberwithin{equation}{section}

%%%%%%%%%%%%%%%%%%%%%%%%%%%%%%%%%%%%%%%%%%%%%%%%%%%%%%%%%%%%%%%%%%%%%%%%%%%%%%
\section{Two generation mixing in~\CP~transformations}
%%%%%%%%%%%%%%%%%%%%%%%%%%%%%%%%%%%%%%%%%%%%%%%%%%%%%%%%%%%%%%%%%%%%%%%%%%%%%%
\label{app-mixings}

In this appendix we discuss the solutions of \eqref{4.22}.
Every matrix $U\in U(2)$ can be represented as
\begin{equation}\label{A.1}
U = e^{i\xi}
  \begin{pmatrix} \alpha & \beta \\ -\beta^\ast & \alpha^\ast \end{pmatrix}
\end{equation}
with $\xi$ real and 
\begin{equation}\label{A.2}
\abs{\alpha}^2 + \abs{\beta}^2 = 1\,.
\end{equation}
Inserting \eqref{A.1} in \eqref{4.22} we get
\begin{align}\label{A.3}
\abs{\alpha}^2 - \beta^2 &= - e^{i \varkappa}\,,
\notag\\
\alpha(\beta+\beta^\ast) &= 0\,,
\notag\\
\abs{\alpha}^2 - \beta^{\ast\,2} &= -e^{i\varkappa}\,.
\end{align}
It follows that
\begin{align}
\label{A.4}
e^{-i \varkappa} &= e^{i \varkappa}\,,
\\
\label{A.5}
\Rightarrow\quad e^{i\varkappa} &= \pm 1\,.
\end{align}
Thus we have two different types of solutions of \eqref{4.22}: type~(a) where 
$e^{i\varkappa}=1$ and type~(b) where $e^{i\varkappa}=-1$.

For the case~(a) we find from \eqref{A.3}
\begin{align}
\label{A.6}
\abs{\alpha}^2-\beta^2 &= -1\,,\notag\\
\beta^2 &= \beta^{\ast\,2}\\
\intertext{and together with \eqref{A.2} we have}
\label{A.7}
\abs{\beta}^2 + \beta^2 &= 2\,.
\intertext{This gives}
\label{A.8}
\beta^2 &= 1\,, \notag\\
\beta &= \pm 1\,, \notag\\
\alpha &= 0\,.
\end{align}
Thus, we get for the case~(a) the solution of \eqref{4.22} for $\beta=1$ as 
\begin{equation}\label{A.9}
U^{(a)} = e^{i\xi} \begin{pmatrix} 0 & 1 \\ -1 & 0 \end{pmatrix}\,.
\end{equation}
Taking $\beta=-1$ gives the same after a redefinition of the phase $\xi$.
This proves \eqref{4.23}.

Turning to the case~(b), $e^{i\varkappa}=-1$, we get from \eqref{A.3}
\begin{align}\label{A.10}
\abs{\alpha}^2-\beta^2 &= 1\,, \notag\\
\alpha(\beta+\beta^\ast) &= 0\,.
\end{align}
Together with \eqref{A.2} we have
\begin{align}
\label{A.11}
\beta^2 &= -\abs{\beta}^2\,, \notag\\
\beta &= \pm \,i\,\beta' \qquad\text{with } \beta'\geq 0\,.
\end{align}
Defining
\begin{equation}\label{A.12}
\alpha' = \mp\,i\,\alpha
\end{equation}
and inserting in \eqref{A.1} we get
\begin{align}\label{A.13}
U^{(b)} &= (\pm i) e^{i\xi}
  \begin{pmatrix} \alpha' & \beta'\\ \beta' & -\alpha'^{\,\ast} \end{pmatrix}\;,\notag\\
\text{with~}&\beta' \geq 0\,, \quad \abs{\alpha'}^2 + \beta'^2 = 1\,.
\end{align}
A redefinition of the phase $\xi$ and the variables $\alpha',\beta'$ proves 
\eqref{4.24}.

%%%%%%%%%%%%%%%%%%%%%%%%%%%%%%%%%%%%%%%%%%%%%%%%%%%%%%%%%%%%%%%%%%%%%%%%%%%%%%
\section{Invariant couplings for~two~lepton~families, details}
%%%%%%%%%%%%%%%%%%%%%%%%%%%%%%%%%%%%%%%%%%%%%%%%%%%%%%%%%%%%%%%%%%%%%%%%%%%%%%
\label{app-leptons}

In this appendix we discuss the structure of the lepton--Higgs coupling
\eqref{4.32} requiring invariance under all four transformations
\CPi, \CPa, \CPb\ and \CPc.
We split the discussion according to the different possibilities for the lepton
masses in the following subsections.

%%%%%%%%%%%%%%%%%%%%%%%%%%%%%%%%%%%%%%%%%%%%%%%%%%%%%%%%%%%%%%%%%%%%%%%%%%%%%%
\subsection{The case of different non-vanishing masses}
%\label{app-leptonsb}

In this subsection we discuss the structure of the lepton--Higgs coupling 
requiring invariance under all four transformations
\CPi, \CPa, \CPb\ and \CPc\ for the case
\begin{equation}\label{B.1}
c^{(1)}_{l\,2} > 0\,,\quad c^{(1)}_{l\,3} > 0\,,\quad
c^{(1)}_{l\,2} \neq c^{(1)}_{l\,3}\,;
\end{equation}
see \eqref{4.34}.
We have to see if matrices $U^{(l)}_R$ and $U^{(l)}_L$ of type (a) (see 
\eqref{4.23}) or type (b) (see \eqref{4.24}) can be found such that 
\eqref{4.33} can be satisfied.

Let us first note that the diagonal form for $C^{(1)}_l$ \eqref{4.11}
still allows one to redefine $l_{\alpha R}$ and 
$(\nu_{\alpha L},l_{\alpha L})^\text{T}$ for given $\alpha$ by multiplication with an
arbitrary phase factor.
If $C^{(2)}_{l23}\neq 0$, we can use this to require, without loss of 
generality, 
\begin{equation}\label{B.2}
C^{(2)}_{l\,23} > 0\,.
\end{equation}
Alternatively, if $C^{(2)}_{l32}\neq 0$, we can use the above freedom of phase 
factors to require without loss of generality
\begin{equation}\label{B.2a}
C^{(2)}_{l\,32} > 0\,.
\end{equation}

\subsubsection{The symmetry \CPb}
\label{subsec-lep2}

Now we impose \CPb\ symmetry.
With $W_{ji}=\delta_{ji}$ (see Tab.~\ref{tab-cphiggs}) we get from \eqref{4.33}
\begin{align}
\label{B.3}
U_R^{(l)\,\trans}\, C_l^{(1)\,\ast}\, U_L^{(l)\,\ast} &= C_l^{(1)}\,,
\\
U_R^{(l)\,\trans}\, C_l^{(2)\,\ast}\, U_L^{(l)\,\ast} &= C_l^{(2)}\,.
\label{B.4}
\end{align}
From \eqref{4.11} we have
\begin{equation}\label{B.5}
C_l^{(1)} = C_l^{(1)\,\ast} = C_l^{(1)\,\trans}\,.
\end{equation}
Together with \eqref{B.3} we get 
\begin{equation}\label{B.6}
( U_R^{(l)}\,U_R^{(l)\,\ast} )^\trans\, C_l^{(1)}\, (U_L^{(l)\,\ast} U_L^{(l)} ) =
  C_l^{(1)}\,.
\end{equation}
From \eqref{B.6}, \eqref{4.23} and \eqref{4.24} we see that we have only the 
possibilities
\begin{enumerate}
\item[(I)] $U_R^{(l)}$ and $U_L^{(l)}$ both of type (a), or
\item[(II)] $U_R^{(l)}$ and $U_L^{(l)}$ both of type (b).
\end{enumerate}

Consider first (I), then we have the ansatz (see \eqref{4.23})
\begin{align}\label{B.7}
U_R^{(l)} &= e^{i\xi_R} \epsilon\,,
\notag\\
U_L^{(l)} &= e^{i\xi_L} \epsilon\,,
\end{align}
and \eqref{B.3} gives 
\begin{equation}\label{B.8}
U_R^{(l)\,\trans}\,C^{(1)}_l = -C^{(1)}_l\,U_L^{(l)}\,.
\end{equation}
Inserting \eqref{B.7} in \eqref{B.8} gives 
\begin{align}\label{B.9}
e^{i\xi_R} c^{(1)}_{l\,2} &= e^{i\xi_L} c^{(1)}_{l\,3}\,,
\notag\\
e^{i\xi_R} c^{(1)}_{l\,3} &= e^{i\xi_L} c^{(1)}_{l\,2}\,,
\end{align}
which is not possible if \eqref{B.1} holds.

Thus we are left with possibility (II), and we can make the ansatz (see 
\eqref{4.24})
\begin{align}\label{B.10}
U_R^{(l)} &= e^{i\xi_R}
  \begin{pmatrix} \alpha_R & \beta_R \\ \beta_R & -\alpha_R^\ast \end{pmatrix},
\notag\\
U_L^{(l)} &= e^{i\xi_L}
  \begin{pmatrix} \alpha_L & \beta_L \\ \beta_L & -\alpha_L^\ast \end{pmatrix},
\end{align}
where
\begin{align}\label{B.11}
\beta_R \geq 0\,, \quad & \beta_L \geq 0\,,
\notag\\
\abs{\alpha_R}^2 + \beta_R^2 &= 1\,,
\notag\\
\abs{\alpha_L}^2 + \beta_L^2 &= 1\,.
\end{align}
From \eqref{B.3} we get now
\begin{equation}\label{B.12}
U_R^{(l)\,\trans}\,C^{(1)}_l = C^{(1)}_l\,U_L^{(l)} \;,
\end{equation}
which gives
\begin{align}\label{B.13}
&e^{i\xi_R}
\begin{pmatrix} \alpha_R c^{(1)}_{l\,2} & \beta_R c^{(1)}_{l\,3} \\
                \beta_R c^{(1)}_{l\,2}  & -\alpha_R^\ast c^{(1)}_{l\,3} 
\end{pmatrix}
=\notag\\
&e^{i\xi_L}
\begin{pmatrix} c^{(1)}_{l\,2}\alpha_L  & c^{(1)}_{l\,2} \beta_L \\
                c^{(1)}_{l\,3}\beta_L  & - c^{(1)}_{l\,3} \alpha_L^\ast 
\end{pmatrix}\,.
\end{align}
With \eqref{B.1} we see that the equality of the diagonal matrix elements in 
\eqref{B.13} can only be fulfilled if
\begin{equation}\label{B.14}
\abs{\alpha_R} = \abs{\alpha_L}\,.
\end{equation}
But this implies from \eqref{B.11}
\begin{equation}\label{B.15}
\beta_R = \beta_L\,,
\end{equation}
and looking now at the off-diagonal matrix elements in \eqref{B.13} we get, due 
to $c^{(1)}_{l2}\neq c^{(1)}_{l3}$, 
\begin{equation}\label{B.16}
\beta_R = \beta_L = 0,
\end{equation}
which implies $|\alpha_R|=|\alpha_L|=1$.
Inserting this in \eqref{B.10} we see the \eqref{B.12} can only be solved if 
\begin{equation}\label{B.17}
U_R^{(l)} = U_L^{(l)} =
  \begin{pmatrix} e^{i\xi_2} & 0 \\ 0 & e^{i\xi_3} \end{pmatrix}\,.
\end{equation}

Turning now to $C^{(2)}_l$ we find from \eqref{B.4}
\begin{equation}\label{B.18}
U_R^{(l)\,\trans}\, C^{(2)\,\ast}_l = C^{(2)}_l\, U_L^{(l)}\,.
\end{equation}
Inserting here \eqref{B.17} we find 
\begin{align}\label{B.19}
&\begin{pmatrix}
  e^{i\xi_2} C^{(2)\,\ast}_{l\,22} & e^{i\xi_2} C^{(2)\,\ast}_{l\,23} \\
  e^{i\xi_3} C^{(2)\,\ast}_{l\,32} & e^{i\xi_3} C^{(2)\,\ast}_{l\,33}
\end{pmatrix}
=\notag\\
&\begin{pmatrix}
  C^{(2)}_{l\,22} e^{i\xi_2} & C^{(2)}_{l\,23} e^{i\xi_3} \\
  C^{(2)}_{l\,32} e^{i\xi_2} & C^{(2)}_{l\,33} e^{i\xi_3}
\end{pmatrix}\,.
\end{align}
From \eqref{B.19} we see immediately that we must have
\begin{align}\label{B.20}
C^{(2)\,\ast}_{l\,22} &= C^{(2)}_{l\,22}\,,
\notag\\
C^{(2)\,\ast}_{l\,33} &= C^{(2)}_{l\,33}\,.
\end{align}
If $C^{(2)}_{l23}\neq 0$ we can impose $C^{(2)}_{l23}>0$; see \eqref{B.2}.
Then \eqref{B.19} can only be fulfilled for $e^{i\xi_2}=e^{i\xi_3}$ and this 
implies
\begin{equation}\label{B.21}
C^{(2)\,\ast}_{l\,32} = C^{(2)}_{l\,32}\,.
\end{equation}
If $C^{(2)}_{l32}\neq 0$ we can use \eqref{B.2a} and come to the conclusion 
\begin{equation}\label{B.22}
C^{(2)\,\ast}_{l\,23} = C^{(2)}_{l\,23}\,.
\end{equation}

To summarise: we find as necessary condition for the leptonic Yukawa coupling 
\eqref{4.32} to allow for a \CPb\ symmetry that $C^{(2)}_l$ is a real matrix,
\begin{equation}\label{B.23}
C^{(2)\,\ast}_l = C^{(2)}_l\,.
\end{equation}
It is easy to see that \eqref{B.23} is also sufficient for \CPb\ invariance.
We just have to set $\xi_2=\xi_3$ in \eqref{B.17} and \eqref{B.19}.

In the following the \CPb\ symmetry condition \eqref{B.23} 
will be supposed to hold.

\subsubsection{The symmetry \CPa}
%\label{subsec-lep1}
Now we impose in addition to \CPb\ the symmetry \CPa.
That is, we look for solutions of \eqref{4.33} with $W_{ji}=\sigma^3_{ji}\,$;
see Tab.~\ref{tab-cphiggs}.
We get then the conditions
\begin{align}
\label{B.24}
U_R^{(l)\,\trans}\, C^{(1)}_l\, U_L^{(l)\,\ast} &= C^{(1)}_l\,,
\\
\label{B.25}
-U_R^{(l)\,\trans}\, C^{(2)}_l\, U_L^{(l)\,\ast} &= C^{(2)}_l\,,
\end{align}
where we already used $C^{(j)}_l=C^{(j)*}_l,~j=1,2$.
Note that with \eqref{B.5}, \eqref{B.3} is equivalent to \eqref{B.24}.
Therefore, we can take over all the results from~\eqref{B.3} up to 
\eqref{B.17}.
Also here $U_R^{(l)}$ and $U_L^{(l)}$ must be both of type (a) or both of type (b).
Only type (b) is possible and we must have 
\begin{equation}\label{B.26}
U_R^{(l)} = U_L^{(l)} =
  \begin{pmatrix} e^{i\xi_2} & 0 \\ 0 & e^{i\xi_3} \end{pmatrix}\,.
\end{equation}
From \eqref{B.25} we get now
\begin{equation}
\label{B.27}
-U_R^{(l)\,\trans}\, C^{(2)}_l = C^{(2)}_l\, U_L^{(l)}\,,
\end{equation}
from which follows that
\begin{align}\label{B.28}
-&\begin{pmatrix}
  e^{i\xi_2} C^{(2)}_{l\,22} & e^{i\xi_2} C^{(2)}_{l\,23} \\
  e^{i\xi_3} C^{(2)}_{l\,32} & e^{i\xi_3} C^{(2)}_{l\,33}
\end{pmatrix}=\notag\\
&
\begin{pmatrix}
  C^{(2)}_{l\,22} e^{i\xi_2} & C^{(2)}_{l\,23} e^{i\xi_3} \\
  C^{(2)}_{l\,32} e^{i\xi_2} & C^{(2)}_{l\,33} e^{i\xi_3}
\end{pmatrix}\,.
\end{align}
This gives immediately as a necessary condition for \eqref{B.28} to be solvable
\begin{align}\label{B.29}
C^{(2)}_{l\,22} &= 0\,,
\notag\\
C^{(2)}_{l\,33} &= 0\,.
\end{align}
Setting
\begin{equation}\label{B.30}
e^{i\xi_2} = -e^{i\xi_3}
\end{equation}
in \eqref{B.26} and \eqref{B.28}, we see that \eqref{B.29} is also sufficient 
for \eqref{B.27} and thus \eqref{B.24} and \eqref{B.25} to have a solution.

The result of this section is that \CPb\ and \CPa\ together can be implemented
as symmetries of $\mathscr{L}_{\mathrm{Yuk},l}$ if and only if $C^{(2)}_l$ is
off-diagonal:
\begin{equation}\label{B.31}
C^{(2)}_l =
  \begin{pmatrix} 0 & C^{(2)}_{l\,23} \\ C^{(2)}_{l\,32} & 0 \end{pmatrix}
\end{equation}
and real.
We shall impose this condition in the following.

\subsubsection{The symmetry \CPi}
\label{subsec-lepi}

From \eqref{4.33} we get here with $W=\epsilon$ (see
Tab.~\ref{tab-cphiggs}) and 
$C^{(j)*}_l=C^{(j)}_l$
\begin{align}
\label{B.32}
U_R^{(l)\,\trans}\, C^{(1)}_l\, U_L^{(l)\,\ast} &= C^{(2)}_l\,,
\\
\label{B.33}
U_R^{(l)\,\trans}\, C^{(2)}_l\, U_L^{(l)\,\ast} &= -C^{(1)}_l\,.
\end{align}
From these we get 
\begin{equation}\label{B.34}
( U_R^{(l)}\,U_R^{(l)\,\ast} )^\trans\, C^{(1)}_l\, (U_L^{(l)\,\ast}\,U_L^{(l)})
= -C^{(1)}_l\,.
\end{equation}
Furthermore, we find from \eqref{B.32}
\begin{equation}\label{B.34a}
\abs{\det C^{(2)}_l} = \abs{\det C^{(1)}_l}\,,
\end{equation}
from which we get
\begin{equation}
\abs{C^{(2)}_{l\,23}\,C^{(2)}_{l\,32}}= c^{(1)}_{l\,2}\,c^{(1)}_{l\,3}\neq 0\,.
\end{equation}
Therefore we have $C^{(2)}_{l23}\neq 0$ and shall use in the following the phase 
choice $C^{(2)}_{l23}>0$; see \eqref{B.2}.

Looking now at \eqref{4.23} and \eqref{4.24} we see that \eqref{B.34} can only 
be satisfied if 
\begin{enumerate}
\item[(I)] $U_R^{(l)}$ is of type (a) and $U_L^{(l)}$ is of type (b), or 
\item[(II)] $U_R^{(l)}$ is of type (b) and $U_L^{(l)}$ is of type (a).
\end{enumerate}

We start with case~(I), where we have to set 
\begin{align}
\label{B.35}
U_R^{(l)} &= e^{i\xi_R} \begin{pmatrix} 0 & 1 \\ -1 & 0 \end{pmatrix}\,,
\\
\label{B.36}
U_L^{(l)} &= e^{i\xi_L}
  \begin{pmatrix} \alpha & \beta \\ \beta & -\alpha^\ast \end{pmatrix}\,,
  \quad
  \beta \ge 0\,,\; \abs{\alpha}^2 + \beta^2 = 1\,.
\end{align}
From \eqref{B.32} and \eqref{B.33} we find 
\begin{equation}
\label{B.37}
U_R^{(l)\,\trans}\, C^{(1)}_l = C^{(2)}_l\, U_L^{(l)}\,,
\end{equation}
from which follows that
\begin{equation}
\label{B.38}
e^{i\xi_R}
  \begin{pmatrix} 0 & -c^{(1)}_{l\,3} \\ c^{(1)}_{l\,2} & 0 \end{pmatrix}
=
e^{i\xi_L}
\begin{pmatrix}
  C^{(2)}_{l\,23}\, \beta & -C^{(2)}_{l\,23}\, \alpha^\ast \\
  C^{(2)}_{l\,32}\, \alpha & C^{(2)}_{l\,32}\, \beta
\end{pmatrix}\,.
\end{equation}
From \eqref{B.34a} and \eqref{B.38} we see that we must have
\begin{gather}
\label{B.39}
\beta = 0\,,\qquad \abs{\alpha} = 1\,,
\\
\label{B.40}
C^{(2)}_{l\,23} = e^{i(\xi_R-\xi_L)}\alpha\,c^{(1)}_{l\,3}\,,
\\
\label{B.41}
C^{(2)}_{l\,32} = e^{i(\xi_R-\xi_L)}\alpha^\ast\,c^{(1)}_{l\,2}\,.
\end{gather}
With \eqref{B.1} and \eqref{B.2} \eqref{B.40} implies 
\begin{equation}\label{B.42}
e^{i(\xi_R-\xi_L)} \alpha = 1\,.
\end{equation}
Since also $C^{(2)}_{l32}$ must be real; see \eqref{B.23}, we get from 
\eqref{B.41} that also 
\begin{equation}\label{B.43}
e^{i(\xi_R-\xi_L)} \alpha^\ast = \pm 1
\end{equation}
must hold.
There are four solutions to \eqref{B.42} and \eqref{B.43}:
\begin{subequations}\label{B.44}
\begin{alignat}{3}
\label{B.44a}
\alpha &= \phantom{-}\alpha^\ast &&= e^{i(\xi_L-\xi_R)} &&= +1\,, \\
\label{B.44b}
\alpha &= \phantom{-}\alpha^\ast &&= e^{i(\xi_L-\xi_R)} &&= -1\,, \\
\label{B.44c}
\alpha &= -\alpha^\ast &&= e^{i(\xi_L-\xi_R)} &&= +i\,, \\
\label{B.44d}
\alpha &= -\alpha^\ast &&= e^{i(\xi_L-\xi_R)} &&= -i\,.
\end{alignat}
\end{subequations}
The corresponding solutions of \eqref{B.38} are as follows.
From \eqref{B.44a} and \eqref{B.44b} we get
\begin{align}\label{B.45a}
C^{(2)}_l & =
  \begin{pmatrix} 0 & c^{(1)}_{l\,3}\\ c^{(1)}_{l\,2} & 0 \end{pmatrix}\,,
\notag\\
U_R^{(l)} &= e^{i\xi_R} \begin{pmatrix} 0 & 1 \\ -1 & 0 \end{pmatrix}\,,
\notag\\
U_L^{(l)} &= e^{i\xi_R} \begin{pmatrix} 1 & 0 \\ 0 & -1 \end{pmatrix}\,.
\end{align}
From \eqref{B.44c} and \eqref{B.44d} we get
\begin{align}\label{B.45b}
C^{(2)}_l & =
  \begin{pmatrix} 0 & c^{(1)}_{l\,3}\\ -c^{(1)}_{l\,2} & 0 \end{pmatrix}\,,
\notag\\
U_R^{(l)} &= e^{i\xi_R} \begin{pmatrix} 0 & 1 \\ -1 & 0 \end{pmatrix}\,,
\notag\\
U_L^{(l)} &= e^{i\xi_R} \begin{pmatrix} -1 & 0 \\ 0 & -1 \end{pmatrix}\,.
\end{align}

Turning to case~(II) we have the ansatz
\begin{align}
\label{B.45c}
U_R^{(l)} &= e^{i\xi_R}
  \begin{pmatrix} \alpha & \beta \\ \beta & -\alpha^\ast \end{pmatrix}\,,
  \quad
\beta \ge 0\,,\; \abs{\alpha}^2 + \beta^2 = 1\,,
\\
\label{B.45d}
U_L^{(l)} &= e^{i\xi_L}
  \begin{pmatrix} 0 & 1 \\ -1 & 0 \end{pmatrix}\,.
\end{align}
From \eqref{B.32} and \eqref{B.33} we get
\begin{equation}\label{B.46}
C^{(1)}_l\, U_L^{(l)} = U_R^{(l)\,\trans}\, C^{(2)}_l\,,
\end{equation}
from which we find
\begin{equation}\label{B.47}
e^{i\xi_L}
  \begin{pmatrix} 0 & c^{(1)}_{l\,2} \\ -c^{(1)}_{l\,3} & 0 \end{pmatrix}
= e^{i\xi_R}
  \begin{pmatrix} \beta\, C^{(2)}_{l\,32} & \alpha\, C^{(2)}_{l\,23} \\
                  -\alpha^\ast\, C^{(2)}_{l\,32} & \beta\, C^{(2)}_{l\,23}
  \end{pmatrix}\,.
\end{equation}
Using \eqref{B.34a} and \eqref{B.2} we find from \eqref{B.47}
\begin{gather}
\label{B.48}
\beta = 0\,,\qquad \abs{\alpha} = 1\,;
\\
C^{(2)}_{l\,23} = e^{i(\xi_L-\xi_R)}\alpha^\ast\,c^{(1)}_{l\,2}\,,
 \notag\\
\label{B.49}
C^{(2)}_{l\,32} = e^{i(\xi_L-\xi_R)}\alpha\,c^{(1)}_{l\,3}\,.
\end{gather}
Complex conjugation in \eqref{B.49} gives
\begin{align}\label{B.50}
C^{(2)}_{l\,23} &= e^{i(\xi_R-\xi_L)}\alpha\,c^{(1)}_{l\,2}\,,
\notag\\
C^{(2)}_{l\,32} &= e^{i(\xi_R-\xi_L)}\alpha^\ast\,c^{(1)}_{l\,3}\,.
\end{align}
Apart from the exchange of $c^{(1)}_{l2}$ and $c^{(1)}_{l3}$ \eqref{B.50} is 
identical to \eqref{B.40} and \eqref{B.41}.
Thus we find that \eqref{B.50} only has solutions if $\alpha$ and 
$\exp\big[i(\xi_L-\xi_R)]$ are equal to one of the four cases shown in 
\eqref{B.44}.

From \eqref{B.44a} and \eqref{B.44b} we get here
\begin{align}\label{B.51}
C^{(2)}_l & =
  \begin{pmatrix} 0 & c^{(1)}_{l\,2}\\ c^{(1)}_{l\,3} & 0 \end{pmatrix}\,,
\notag\\
U_R^{(l)} &= e^{i\xi_L} \begin{pmatrix} 1 & 0 \\ 0 & -1 \end{pmatrix}\,,
\notag\\
U_L^{(l)} &= e^{i\xi_L} \begin{pmatrix} 0 & 1 \\ -1 & 0 \end{pmatrix}\,.
\end{align}
From \eqref{B.44c} and \eqref{B.44d} we get here
\begin{align}\label{B.52}
C^{(2)}_l & =
  \begin{pmatrix} 0 & c^{(1)}_{l\,2}\\ -c^{(1)}_{l\,3} & 0 \end{pmatrix}\,,
\notag\\
U_R^{(l)} &= e^{i\xi_L} \begin{pmatrix} 1 & 0 \\ 0 & 1 \end{pmatrix}\,,
\notag\\
U_L^{(l)} &= e^{i\xi_L} \begin{pmatrix} 0 & 1 \\ -1 & 0 \end{pmatrix}\,.
\end{align}

We see that requiring the symmetries \CPb, \CPa\ and \CPi\ to hold leads to
only four distinct possibilities for $C^{(2)}_l$ as shown in \eqref{B.45a},
\eqref{B.45b}, \eqref{B.51} and \eqref{B.52} and Tab.~\ref{tab-couplppneq}.
Of course, all this applies only under the condition that \eqref{4.34} holds.
It remains to be seen if these four cases also allow for the implementation of the 
\CPc\ symmetry.

\subsubsection{The symmetry \CPc}
%\label{subsec-lep3}

Here we have to look for solutions of \eqref{4.33} setting 
$W_{ji}=\sigma^1_{ji}$; see Tab.~\ref{tab-cphiggs}.
This gives with $C^{(j)*}_l=C^{(j)}_l$
\begin{align}\label{B.53}
U_R^{(l)\,\trans}\, C^{(1)}_l\, U_L^{(l)\,\ast} &= C^{(2)}_l\,,
\notag\\
U_R^{(l)\,\trans}\, C^{(2)}_l\, U_L^{(l)\,\ast} &= C^{(1)}_l\,.
\end{align}
From this we get 
\begin{equation}\label{B.54}
( U_R^{(l)}\,U_R^{(l)\,\ast} )^\trans\, C_l^{(1)}\, (U_L^{(l)\,\ast} U_L^{(l)} ) =
  C_l^{(1)}\,,
\end{equation}
which shows that both, $U_R^{(l)}$ and $U_L^{(l)}$, have to be of type (a) as in 
\eqref{4.23} or of type (b) as in \eqref{4.24}.
We have already found that $C^{(2)}_l$ can be only as in \eqref{B.45a}, 
\eqref{B.45b}, \eqref{B.51} or \eqref{B.52}.
It is easy to see that none of these cases allows for a solution of \eqref{B.53}
if $U_R^{(l)}$ and $U_L^{(l)}$ are of type (a).
If both $U_R^{(l)}$ and $U_L^{(l)}$ are of type (b) we have solutions in all four 
cases.
For $C^{(2)}_l$ as in \eqref{B.45a} we get a solution of \eqref{B.53}
setting
\begin{align}\label{B.55}
U_R^{(l)} &= \sigma^1\,,
\notag\\
U_L^{(l)} &= \unitmatrix_2\,.
\end{align}
For $C^{(2)}_l$ as in \eqref{B.45b} we set 
\begin{align}\label{B.56}
U_R^{(l)} &= \sigma^1\,,
\notag\\
U_L^{(l)} &= -\sigma^3\,.
\end{align}
For $C^{(2)}_l$ as in \eqref{B.51} we set
\begin{align}\label{B.57}
U_R^{(l)} &= \unitmatrix_2\,,
\notag\\
U_L^{(l)} &= \sigma^1\,.
\end{align}
For $C^{(2)}_l$ as in \eqref{B.52} we set
\begin{align}\label{B.58}
U_R^{(l)} &= \sigma^3\,,
\notag\\
U_L^{(l)} &= \sigma^1\,.
\end{align}

This completes the proof that for all cases of $C^{(2)}_l$ listed in 
Tab.~\ref{tab-couplppneq} we can implement all symmetries 
\CPa, \CPb, \CPc\ and \CPi.
As we have shown, no other form of $C^{(2)}_l$ allows this 
to happen if \eqref{4.34} 
holds for $C^{(1)}_l$.

%%%%%%%%%%%%%%%%%%%%%%%%%%%%%%%%%%%%%%%%%%%%%%%%%%%%%%%%%%%%%%%%%%%%%%%%%%%%%%
\subsection{The equal mass case}
%\label{subsec-lepequal}

Here we suppose
\begin{equation}\label{C.1}
c^{(1)}_{l\,2} = c^{(1)}_{l\,3} > 0
\end{equation}
to hold.
We have then
\begin{equation}\label{C.2}
C^{(1)}_l = c^{(1)}_{l\,2}\,\unitmatrix_2
\end{equation}
in \eqref{4.32}.
We want to find the general structure of $C^{(2)}_l$ compatible with invariance 
under all four \CP\ transformations.

We start with the \CPi\ symmetry which requires \eqref{4.33} to hold 
with $W_{ji}=\epsilon_{ji}$; see Tab.~\ref{tab-cphiggs}.
This leads to 
\begin{align}
\label{C.3}
c^{(1)}_{l\,2}\, U_R^{(l)\,\trans}\, U_L^{(l)\,\ast} &= C^{(2)}_l\,,
\\
\label{C.4}
U_R^{(l)\,\trans}\, C^{(2)\,\ast}_l\, U_L^{(l)\,\ast} &=
  -c^{(1)}_{l\,2}\,\unitmatrix_2\,.
\end{align}
From \eqref{C.3} we see that $C^{(2)}_l/c^{(1)}_{l2}$ is a unitary matrix.
Therefore we can make a basis change of $l_{\alpha R}$ and
$(\nu_{\alpha L},l_{\alpha L})^{\text T}$ in order to diagonalise $C^{(2)}_l$.
Indeed, consider the basis change
\begin{align}\label{C.5}
l_{\alpha\,R} &\rightarrow V_{\alpha\beta}\,l_{\beta\,R}\,,
\notag\\
\begin{pmatrix} \nu_{\alpha\,L} \\ l_{\alpha\,L} \end{pmatrix}
  &\rightarrow V_{\alpha\beta} 
  \begin{pmatrix} \nu_{\beta\,L} \\ l_{\beta\,L} \end{pmatrix}\,,
\end{align}
with $V \in U(2)$:
\begin{align}\label{C.6}
V &= ( V_{\alpha\beta} )\,, &
V\,V^\dagger &= V^\dagger\,V = \unitmatrix_2\,.
\end{align}
This leads to 
\begin{equation}\label{C.7}
C^{(j)}_l \rightarrow V^\dagger\, C^{(j)}_l\, V\,,\qquad j=1,2\,.
\end{equation}
With a suitable $V$ we can achieve 
\begin{equation}\label{C.8}
C^{(2)}_l = c^{(1)}_{l\,2}
  \begin{pmatrix} e^{i\eta_2} & 0 \\ 0 & e^{i\eta_3} \end{pmatrix}\,,
\end{equation}
with $\eta_2$ and $\eta_3$ real.
Note that the transformation \eqref{C.7} does not affect $C^{(1)}_l$ of 
\eqref{C.2}.

Taking the complex conjugate of \eqref{C.4} and inserting $C^{(2)}_l$ from 
\eqref{C.3} we get
\begin{equation}\label{C.9}
(U_R^{(l)}\,U_R^{(l)\,\ast} )^\trans\,(U_L^{(l)\,\ast}\,U_L^{(l)})
  = -\unitmatrix_2\,.
\end{equation}
This shows that there are only two possibilities:
\begin{itemize}
\item[(I)] $U_R^{(l)}$ of type (a); see \eqref{4.23}, and $U_L^{(l)}$ of type (b),
see \eqref{4.24}, or
\item[(II)] $U_R^{(l)}$ of type (b) and $U_L^{(l)}$ of type (a).
\end{itemize}
We start by considering case~(I) where we have the ansatz
\begin{align}\label{C.10}
U_R^{(l)} &= e^{i\xi_R} \begin{pmatrix} 0 & 1 \\ -1 & 0 \end{pmatrix}\,,
\notag\\
U_L^{(l)} &= e^{i\xi_L}
  \begin{pmatrix} \alpha & \beta \\ \beta & -\alpha^\ast \end{pmatrix}\,,
\quad \beta \geq 0\,,\; \abs{\alpha}^2 + \beta^2 = 1\,.
\end{align}
From \eqref{C.3} and \eqref{C.8} we get then
\begin{align}
\label{C.11}
c^{(1)}_{l\,2}\, U_R^{(l)\,\trans} &= C^{(2)}_l\, U_L^{(l)}\,,
\\
\label{C.12}
e^{i\xi_R}\begin{pmatrix} 0 & -1 \\ 1 & 0 \end{pmatrix} &=
e^{i\xi_L}
  \begin{pmatrix} e^{i\eta_2}\alpha & e^{i\eta_2}\beta \\
                  e^{i\eta_3}\beta  & -e^{i\eta_3}\alpha^\ast \end{pmatrix}\,.
\end{align}
It follows that
\begin{gather}
\label{C.13}
\alpha = 0\,,\quad \beta=1\,,
\\
-e^{i\xi_R} = e^{i\xi_L} e^{i\eta_2}\,,
\notag\\
\phantom{-}e^{i\xi_R} = e^{i\xi_L} e^{i\eta_3}\,,
\label{C.14}
\\
\label{C.15}
\Rightarrow\quad e^{i(\xi_R-\xi_L)} = e^{i\eta_3} = - e^{i\eta_2}\,.
\end{gather}
Thus we find here the following solution of \eqref{C.3} and \eqref{C.4}:
\begin{align}
\label{C.16}
C^{(2)}_l &= c^{(1)}_{l\,2}\,e^{i\eta_2}
  \begin{pmatrix} 1 & 0 \\ 0 & -1 \end{pmatrix}
\notag\\
 &= c^{(1)}_{l\,2}\,e^{i\eta_2}\sigma^3\,,
\\[2mm]
\label{C.17}
U_R^{(l)} &= e^{i\xi_R}\,\epsilon\,,
\notag\\
U_L^{(l)} &= - e^{i(\xi_R-\eta_2)}\,\sigma^1\,.
\end{align}

Turning now to case~(II) we have to make the ansatz
\begin{align}
\label{C.18}
U_R^{(l)} &= e^{i\xi_R}
  \begin{pmatrix} \alpha & \beta \\ \beta & -\alpha^\ast \end{pmatrix}\,,
\quad
\beta \geq 0\,, \; \abs{\alpha}^2+\beta^2=1\,,
\\
\label{C.19}
U_L^{(l)} &= e^{i\xi_L}\,\epsilon\,.
\end{align}
Here we get from \eqref{C.3} and \eqref{C.4}
\begin{equation}
\label{C.20}
c^{(1)}_{l\,2}\, U_R^{(l)\,\trans} = - C^{(2)}_l\, U_L^{(l)}\,,
\end{equation}
from which follows that
\begin{equation}\label{C.21}
e^{i\xi_R}
  \begin{pmatrix} \alpha & \beta \\ \beta & -\alpha^\ast \end{pmatrix} =
e^{i\xi_L}
  \begin{pmatrix} 0 & -e^{i\eta_2} \\ e^{i\eta_3} & 0 \end{pmatrix}\,.
\end{equation}
This can only be fulfilled if
\begin{gather}\label{C.22}
\alpha = 0\,,\quad \beta=1\,,
\notag\\
e^{i(\xi_R-\xi_L)} = e^{i\eta_3} = - e^{i\eta_2} \,.
\end{gather}
Inserting this in \eqref{C.8} we find that $C^{(2)}_l$ must again have the form 
\eqref{C.16}, and then \eqref{C.20} is solved with 
\begin{align}\label{C.23}
U_R^{(l)} &= e^{i\xi_R}\,\sigma^1\,,
\notag\\
U_L^{(l)} &= -e^{i(\xi_R-\eta_2)}\,\epsilon\,.
\end{align}

Thus, \CPi\ invariance now fixes $C^{(2)}_l$ to be of the form 
\eqref{C.16}.
We shall next impose \CPb\ invariance.
From \eqref{4.33} we find then with $W_{ji}=\delta_{ji}$ the conditions
\begin{align}
\label{C.24}
U_R^{(l)\,\trans}\,U_L^{(l)\,\ast} &= \unitmatrix_2\,,
\\
\label{C.25}
U_R^{(l)\,\trans}\, e^{-i\eta_2}\,\sigma^3\,U_L^{(l)\,\ast} &=
  e^{i\eta_2}\,\sigma^3\,.
\end{align}
From \eqref{C.24} we get immediately
\begin{equation}\label{C.26}
U_R^{(l)} = U_L^{(l)}\,.
\end{equation}
Inserting this in \eqref{C.25} we get
\begin{equation}\label{C.27}
\sigma^3 U_L^{(l)} = e^{-2 i \eta_2}\, U_L^{(l)} \sigma^3\,.
\end{equation}
We can have two cases.
\begin{itemize}
\item[(I)] $U_L^{(l)}$ of type (a); see \eqref{4.23},
\end{itemize}
\begin{equation}\label{C.28}
U_L^{(l)} = e^{i\xi_L}\,\epsilon\,.
\end{equation}
Inserting this in \eqref{C.27} gives
\begin{align}\label{C.29}
e^{2i\eta_2} &= -1\,,
\\
\label{C.30}
e^{i\eta_2} &= \pm i\,.
\end{align}
\begin{itemize}
\item[(II)] $U_L^{(l)}$ of type (b); see \eqref{4.24},
\end{itemize}
\begin{align}\label{C.31}
U_L^{(l)}&=e^{i\xi_L}
  \begin{pmatrix} \alpha & \beta \\ \beta & -\alpha^\ast \end{pmatrix}\,,
 \quad
 \beta \geq 0\,,\quad \abs{\alpha}^2 + \beta^2 = 1\,.
\end{align}
Now \eqref{C.27} gives
\begin{equation}
\label{C.32}
\begin{pmatrix} \alpha & \beta \\ -\beta & \alpha^\ast \end{pmatrix}
  = e^{-2i\eta_2}
  \begin{pmatrix} \alpha & -\beta \\ \beta & \alpha^\ast \end{pmatrix}\,,
\end{equation}
from which we get
\begin{align}
\label{C.33}
(1+e^{-2i\eta_2})\beta &= 0\,,\\
\label{C.33a}
(1-e^{-2i\eta_2})\alpha &= 0\,.
\end{align}
The solutions of~\eqref{C.33} and \eqref{C.33a} and are as follows.
For $\alpha\neq 0$ we get
\begin{align}\label{C.34}
e^{-2i\eta_2} &= 1\,,
\notag\\
e^{i\eta_2} &= \pm 1\,,
\notag\\
\beta=0\,, &\quad \abs{\alpha}=1\,.
\end{align}
For $\alpha=0$ we must have $\beta\neq 0$; see \eqref{C.31}, and we get
\begin{align}\label{C.35}
e^{-2i\eta_2} &= -1\,,
\notag\\
e^{i\eta_2} &= \pm i\,,
\notag\\
\alpha=0\,, &\quad \beta=1\,.
\end{align}

In summary: we see from~\eqref{C.16}, \eqref{C.30}, \eqref{C.34} and \eqref{C.35} that 
imposition of the \CPi\ and \CPb\ symmetries requires $C^{(2)}_l$ to be of one 
of the following forms:
\begin{subequations}\label{C.36}
\begin{align}
C^{(2)}_l &= c^{(1)}_{l\,2}\,\sigma^3\,,
\\
C^{(2)}_l &= -c^{(1)}_{l\,2}\,\sigma^3\,,
\\
C^{(2)}_l &= i\,c^{(1)}_{l\,2}\,\sigma^3\,,
\\
C^{(2)}_l &= -i\,c^{(1)}_{l\,2}\,\sigma^3\,.
\end{align}
\end{subequations}
For all these cases we can also implement \CPa\ and \CPc\ invariance
transformations.
That is, we can always find appropriate $U_R^{(l)}$ and $U_L^{(l)}$ solving 
\eqref{4.33} with $W_{ji}=\sigma^3_{ji}$ and $W_{ji}=\sigma^1_{ji}$ for 
\CPa\ and \CPc, respectively; see Tab.~\ref{tab-cphiggs}.
We list the corresponding matrices in Tab.~\ref{tab-lepequalmixings}.

\begin{table}
\begin{tabular}{|c||c||c|c|}
\hline
$C^{(2)}_l/c^{(1)}_{l2}$
  & \CPa & \multicolumn{2}{c|}{\CPc}\\
\hline
  & $\;U_R^{(l)}=U_L^{(l)}\;$& $\;U_R^{(l)}\;$ & $\;U_L^{(l)}\;$\\
\hline
$\phantom{+}\sigma^3$&$\epsilon$&$\phantom{+}\sigma^3$&${\mathbbm 1}_2$\\
$-\sigma^3$&$\sigma^1$&$-\sigma^3$&${\mathbbm 1}_2$\\
$\phantom{+}i\sigma^3$&${\mathbbm 1}_2$&$\phantom{+}i\sigma^3$&${\mathbbm 1}_2$\\
$-i\sigma^3$&${\mathbbm 1}_2$&$-i\sigma^3$&${\mathbbm 1}_2$\\
\hline
\end{tabular}
\caption{\label{tab-lepequalmixings}
Matrices $U_R^{(l)}$ and $U_L^{(l)}$ solving \eqref{4.33} for the allowed 
forms of $C^{(2)}_l$, \eqref{C.36}, for the symmetries \CPa\ and 
\CPc.
}
\end{table}

%%%%%%%%%%%%%%%%%%%%%%%%%%%%%%%%%%%%%%%%%%%%%%%%%%%%%%%%%%%%%%%%%%%%%%%%%%%%%%
\subsection{The massive plus massless case}

Here we discuss the case that we have one massive and one massless lepton; that 
is, we suppose
\begin{equation}\label{C.37}
c^{(1)}_{l\,2} = 0 \,,\qquad c^{(1)}_{l\,3} > 0
\end{equation}
in \eqref{4.11}.
In fact, we shall start here from the more general case
\begin{equation}\label{C.38}
c^{(1)}_{l\,2} \geq 0 \,,\qquad c^{(1)}_{l\,3} > 0
\end{equation}
and prescribe matrices $U^{(l)}_R$ and $U^{(l)}_L$ for the leptons for our four 
\CP\ symmetries as shown in Tab.~\ref{tab-cpzp}.
We require \eqref{4.33} to hold with the matrices of Tab.~\ref{tab-cpzp} for
all four \CPg\ symmetries.

We start with $\CPb = \CPs$.
From \eqref{4.33} and Tab.~\ref{tab-cpzp} we get here 
\begin{align}\label{C.39}
C^{(1)\,\ast}_l &= C^{(1)}_l\,,
\notag\\
C^{(2)\,\ast}_l &= C^{(2)}_l\,.
\end{align}
Thus $C^{(2)}_l$ is constrained to be a real matrix.
Requiring now also \CPa\ invariance we get from \eqref{4.33} and 
Tab.~\ref{tab-cpzp}
\begin{align}
\label{C.40}
\begin{pmatrix} -1 & 0 \\ 0 & 1 \end{pmatrix}
   \begin{pmatrix} c^{(1)}_{l\,2} & 0 \\ 0 & c^{(1)}_{l\,3} \end{pmatrix}
&= \begin{pmatrix} c^{(1)}_{l\,2} & 0 \\ 0 & c^{(1)}_{l\,3} \end{pmatrix}\,,
\\
\label{C.41}
\begin{pmatrix} 1 & 0 \\ 0 & -1 \end{pmatrix}
   \begin{pmatrix} C^{(2)}_{l\,22} & C^{(2)}_{l\,23} \\
                   C^{(2)}_{l\,32} & C^{(2)}_{l\,33} \end{pmatrix}
&= \begin{pmatrix} C^{(2)}_{l\,22} & C^{(2)}_{l\,23} \\
                   C^{(2)}_{l\,32} & C^{(2)}_{l\,33} \end{pmatrix}\,.
\end{align}
It follows that
\begin{align}
\label{C.42}
c^{(1)}_{l\,2} &= 0\,,
\\
\label{C.43}
C^{(2)}_{l\,32} &= C^{(2)}_{l\,33} = 0\,.
\end{align}
Thus, a massless lepton $l_2$ which is implied by \eqref{C.42} is here a
consequence of our ansatz for the \CPa\ symmetry.

The next symmetry to consider is \CPi\ where we get from \eqref{4.33} 
and Tab.~\ref{tab-cpzp}
\begin{equation}\label{C.44}
\begin{pmatrix} 0 & -1 \\ 1 & 0 \end{pmatrix}
\begin{pmatrix} 0 & 0 \\ 0 & c^{(1)}_{l\,3} \end{pmatrix}
\begin{pmatrix} 0 & 1 \\ 1 & 0 \end{pmatrix}
= C^{(2)}_l\,.
\end{equation}
It follows that
\begin{equation}\label{C.45}
C^{(2)}_l =
  \begin{pmatrix} -c^{(1)}_{l\,3} & 0 \\ 0 & 0 \end{pmatrix}\,.
\end{equation}

The remaining relations for the \CPi\ and \CPc\ symmetries 
following from \eqref{4.33} and Tab.~\ref{tab-cpzp} are easily seen to hold
if \eqref{C.42} and \eqref{C.45} are true.

Now we give a general discussion of the case one massless and one massive
lepton.
That is, we suppose \eqref{C.37} to hold and impose invariance under
\CPi, \CPa, \CPb\ and \CPc.

We start with \CPi\ where we look for matrices $U_R^{(l)}$ and $U_L^{(l)}$ such
that
\begin{align}\label{C.46}
U_R^{(l)\,\trans}\, C_l^{(1)\,\ast}\, U_L^{(l)\,\ast} &=  C^{(2)}_l\,, \notag\\
U_R^{(l)\,\trans}\, C_l^{(2)\,\ast}\, U_L^{(l)\,\ast} &= -C^{(1)}_l\,.
\end{align}
See \eqref{4.33} and Tab.~\ref{tab-cphiggs}.
From \eqref{C.46} we get
\begin{equation}\label{C.47}
\left(U_R^{(l)\,\ast}\,U_R^{(l)}\right)^\trans C^{(1)}_l
  \left(U_L^{(l)}\,U_L^{(l)\,\ast}\right) = -C^{(1)}_l\,.
\end{equation}
This shows that we have only two possibilities,
\begin{itemize}
\item[(I)] $U_R^{(l)}$ of type~(a), see \eqref{4.23}, and $U_L^{(l)}$ of
type~(b), see \eqref{4.24}, or
\item[(II)] $U_R^{(l)}$ of type~(b) and $U_L^{(l)}$ of type~(a)\,.
\end{itemize}

For the case~(I) we have the ansatz
\begin{align}\label{C.48}
U_R^{(l)} &= e^{i\xi_R} \begin{pmatrix} 0 & 1\\ -1 & 0 \end{pmatrix}\,,\notag\\
U_L^{(l)} &= e^{i\xi_L}
  \begin{pmatrix} \alpha & \beta \\ \beta & -\alpha^\ast \end{pmatrix}\,,
  \quad
\beta \geq 0\,, \; \abs{\alpha}^2 + \beta^2 = 1\,.
\end{align}
From \eqref{C.46} we get here
\begin{align}\label{C.49}
C_l^{(2)} &= U_R^{(l)\,\trans}\, C_l^{(1)\,\ast}\, U_L^{(l)\,\ast}
\notag\\
&= c^{(1)}_{l\,3}\, e^{i(\xi_R-\xi_L)}
  \begin{pmatrix} -\beta & \alpha \\ 0 & 0 \end{pmatrix}\,.
\end{align}
That is, with $C_l^{(2)}$ from \eqref{C.49} we have \CPi\ invariance.
The phase factor $\exp[i(\xi_R-\xi_L)]$ can be absorbed in the definition of
the field $l_{2\,R}(x)$.
Thus, the only independent solutions are here
\begin{equation}\label{C.50}
C_l^{(2)} = c^{(1)}_{l\,3}
  \begin{pmatrix} -\beta & \alpha \\ 0 & 0 \end{pmatrix}\,.
\end{equation}

For the case~(II) we have the ansatz
\begin{align}\label{C.51}
U_R^{(l)} &= e^{i\xi_R}
  \begin{pmatrix} \alpha & \beta \\ \beta & -\alpha^\ast \end{pmatrix}\,,
  \quad
\beta \geq 0\,, \; \abs{\alpha}^2 + \beta^2 = 1\,,
\notag\\
U_L^{(l)} &= e^{i\xi_L} \begin{pmatrix} 0 & 1 \\ -1 & 0 \end{pmatrix}\,.
\end{align}
From \eqref{C.46} we get here
\begin{equation}\label{C.52}
C^{(2)}_l = c^{(1)}_{l\,3}\, e^{i(\xi_R-\xi_L)}
  \begin{pmatrix} -\beta & 0 \\ \alpha^\ast & 0 \end{pmatrix}\,.
\end{equation}
Absorbing the phase factor $\exp[i(\xi_R-\xi_L)]$ in the definition of the
doublet fields $(\nu_{2\,L}(x),\, l_{2\,L}(x))^\trans$ we get here the
independent solutions as follows:
\begin{equation}\label{C.53}
C^{(2)}_l = c^{(1)}_{l\,3}
  \begin{pmatrix} -\beta & 0 \\ \alpha^\ast & 0 \end{pmatrix}\,.
\end{equation}

Thus we find that \CPi\ symmetry requires $C^{(2)}_l$ to be of the form
\eqref{C.50} or \eqref{C.53}.
Turning now to the symmetries \CPa, \CPb\ and \CPc\ we find that they can
always be implemented for $C^{(2)}_l$ of \eqref{C.50} or \eqref{C.53}.
Thus, these symmetries give no further restrictions for $C^{(2)}_l$.
However, if we require the absence of FCNCs we must have $\alpha = 0$ in
\eqref{C.50} and \eqref{C.53}.
Then we have $\beta=1$, and we are led to the unique form
\begin{equation}\label{C.54}
C^{(2)}_l = c^{(1)}_{l\,3} \begin{pmatrix} -1 & 0 \\ 0 & 0 \end{pmatrix}
\end{equation}
entering in the coupling \eqref{4.42}.

%%%%%%%%%%%%%%%%%%%%%%%%%%%%%%%%%%%%%%%%%%%%%%%%%%%%%%%%%%%%%%%%%%%%%%%%%%%%%%
\section{Invariant couplings for~two~quark~families, details}
%%%%%%%%%%%%%%%%%%%%%%%%%%%%%%%%%%%%%%%%%%%%%%%%%%%%%%%%%%%%%%%%%%%%%%%%%%%%%%
\label{app-quarks}

Here we study the Yukawa coupling $\mathscr{L}_{\mathrm{Yuk},q}$ \eqref{4.49}
supposing
\begin{equation}\label{D.1}
c^{(1)}_{d\,2} \geq 0 \,,\qquad c^{(1)}_{d\,3} > 0
\end{equation}
and prescribing the matrices $U^{(d)}_R$ and $U^{(u)}_L$ for the \CPg\ 
transformations of the quark fields 
(see \eqref{4.17}) 
as follows:
\begin{align}
\label{D.2}
U^{(d)}_R &= U^{(l)}_R\,,
\\
\label{D.2a}
U^{(u)}_L &= U^{(l)}_L\,.
\end{align}
Here $U^{(l)}_R$ and $U^{(l)}_L$ are as in Tab.~\ref{tab-cpzp}.
Note that the ansatz \eqref{D.2} refers to the $d'_{\alpha R}$ fields and 
\eqref{D.2a} to the fields $(u_{\alpha L},d'_{\alpha L})^\text{T}$.

Without loss of generality we may suppose $C^{(1)}_q$ to be of the form 
\eqref{4.13} and \eqref{4.14}.
Imposing now \CPa\ invariance we find from \eqref{4.28} with 
$W_{ji}=\sigma^3_{ji}$ (see Tab.~\ref{tab-cphiggs})
\begin{equation}\label{D.3}
U^{(d)\,\trans}_R\, C^{(j)\,\ast}_q\, U^{(u)\,\ast}_L\, \sigma^3_{ji}
  = C^{(i)}_q\,.
\end{equation}
Inserting here $U^{(d)}_R$ and $U^{(u)}_L$ according to \eqref{D.2}, 
\eqref{D.2a} and Tab.~\ref{tab-cpzp} we get
\begin{equation}
\label{D.3a}
(-\sigma^3)C^{(1)}_q = C^{(1)}_q\,,
\end{equation}
from which follows that
\begin{equation}
\label{D.4}
-\sigma^3\,V
  \begin{pmatrix} c^{(1)}_{d\,2} & 0 \\ 0 & c^{(1)}_{d\,3} \end{pmatrix}
  V^\dagger
 =
  V
  \begin{pmatrix} c^{(1)}_{d\,2} & 0 \\ 0 & c^{(1)}_{d\,3} \end{pmatrix}
  V^\dagger
\end{equation}
and
\begin{equation}
\label{D.5}
  \begin{pmatrix} -V_{22}\,c^{(1)}_{d\,2} & -V_{23}\,c^{(1)}_{d\,3} \\
                   V_{32}\,c^{(1)}_{d\,2} &  V_{33}\,c^{(1)}_{d\,3}
  \end{pmatrix}
  =
  \begin{pmatrix} V_{22}\,c^{(1)}_{d\,2} & V_{23}\,c^{(1)}_{d\,3} \\
                  V_{32}\,c^{(1)}_{d\,2} & V_{33}\,c^{(1)}_{d\,3}
  \end{pmatrix}\,.
\end{equation}
With \eqref{D.1} we get
\begin{equation}\label{D.6}
V_{23} = 0\,,
\end{equation}
which implies from \eqref{4.14}
\begin{equation}\label{D.7}
V_{22}=V_{33}=1\,.
\end{equation}
Inserting this in \eqref{D.5} gives 
\begin{equation}\label{D.8}
c^{(1)}_{d\,2} = 0\,.
\end{equation}
Thus our symmetry requires a massless $d_2$-quark
\begin{equation}\label{D.9}
m_{d\,2} = c^{(1)}_{d\,2}\,\frac{v_0}{\sqrt{2}} = 0\,.
\end{equation}

The analysis of the remaining symmetry requirements runs now along exactly the 
same lines as for the leptons in section~\ref{app-leptons}.
The result is that our principle of maximal \CP\ symmetry with the ansatz 
\eqref{D.2} and \eqref{D.2a} for the matrices $U^{(d)}_R$ and $U^{(u)}_L$ requires
\begin{align}\label{D.10}
C^{(1)}_q &= \begin{pmatrix} 0 & 0 \\ 0 & c^{(1)}_{d\,3} \end{pmatrix}\,,
\notag\\
C^{(2)}_q &= \begin{pmatrix} -c^{(1)}_{d\,3} & 0 \\ 0 & 0 \end{pmatrix}\,.
\end{align}

Finally, we make some remarks on the general analysis for the case of one
massless quark pair $(u_2,~d_2)$ and one massive pair $(u_3,~d_3)$.
That is, we suppose
\begin{align}\label{D.11}
c^{(1)}_{u\,2} &= 0\,, & c^{(1)}_{u\,3} &> 0\,, \notag\\
c^{(1)}_{d\,2} &= 0\,, & c^{(1)}_{d\,3} &> 0\,.
\end{align}
As for the lepton sector, we can show that the principle of maximal
\CP\ invariance together with the requirement of absence of FCNCs leads to the
following structure of the quark--Higgs coupling matrices (see \eqref{C.54}):
\begin{align}
\label{D.12}
C'^{(1)}_q &= \begin{pmatrix} 0 & 0\\ 0 & c^{(1)}_{u\,3} \end{pmatrix}\,,
\notag\\
C'^{(2)}_q &= \begin{pmatrix} -c^{(1)}_{u\,3} & 0\\ 0 & 0 \end{pmatrix}\,,
\\
\label{D.13}
\tilde{C}^{(1)}_q &=\begin{pmatrix} 0 & 0\\ 0 & c^{(1)}_{d\,3} \end{pmatrix}\,,
\notag\\
\tilde{C}^{(2)}_q &=\begin{pmatrix} -c^{(1)}_{d\,3} & 0\\ 0 & 0\end{pmatrix}\,.
\end{align}
Here $\tilde{C}^{(j)}_q$ are the CKM rotated matrices according to
\eqref{4.52}.
From the discussion of the lepton case (see \eqref{C.46}~ff) we see that
\CPi\ invariance is implementable for \eqref{D.12} and \eqref{D.13} only with
certain matrices $U_R^{(u)}$, $U_L^{(u)}$ in 
%\eqref{4.18}
\eqref{4.17}
 and certain
CKM rotated matrices $\tilde{U}_R^{(d)}$, $\tilde{U}_L^{(u)}$ for the
CKM rotated fields in \eqref{4.50}.
Here we have according to \eqref{4.24c}
\begin{align}
\label{D.14}
\tilde{U}_R^{(d)} &= V^\dagger\, U_R^{(d)}\, V^\ast\,,
\\
\label{D.15}
\tilde{U}_L^{(u)} &= V^\dagger\, U_L^{(u)}\, V^\ast\,.
\end{align}
We have for $U_R^{(u)}$, $U_L^{(u)}$ and $\tilde{U}_R^{(d)}$,
$\tilde{U}_L^{(u)}$ only the possibilities (I) and (II) of \eqref{C.48} and
\eqref{C.51}, respectively, with $\alpha=0$ and $\xi_R=\xi_L$.
We have to check for the resulting four cases if we can then fulfil
\eqref{D.15} or, equivalently, with $V=V^\ast$ (see \eqref{4.14})
\begin{equation}\label{D.16}
V\,\tilde{U}_L^{(u)} = U_L^{(u)}\,V\,.
\end{equation}

For both, $U_R^{(u)}$, $U_L^{(u)}$ and $\tilde{U}_R^{(d)}$,
$\tilde{U}_L^{(u)}$, of the type \eqref{C.48} with $\alpha=0$ and $\xi_R=\xi_L$
we have
\begin{align}\label{D.17}
U_R^{(u)} &= e^{i\xi} \epsilon\,,
\notag\\
U_L^{(u)} &= e^{i\xi} \sigma^1\,,
\notag\\
\tilde{U}_R^{(d)} &= e^{i\tilde{\xi}} \epsilon\,,
\notag\\
\tilde{U}_L^{(d)} &= e^{i\tilde{\xi}} \sigma^1\,.
\end{align}

Then \eqref{D.16} can only be fulfilled in two cases.
The first solution is
\begin{align}\label{D.18}
\vartheta &= 0\,, \notag\\ e^{i\tilde{\xi}} &= e^{i\xi}\,,
\end{align}
implying
\begin{align}\label{D.19}
V &= \unitmatrix_2\,, & U_R^{(d)} &= U_R^{(u)}\,.
\end{align}
The second solution is
\begin{align}\label{D.20}
\vartheta &= \pi/2\,, \notag\\ e^{i\tilde{\xi}} &= -e^{i\xi}\,,
\end{align}
implying
\begin{align}\label{D.21}
V &= \epsilon\,, & U_R^{(d)} &= -U_R^{(u)}\,.
\end{align}

For both, $U_R^{(u)}$, $U_L^{(u)}$ and $\tilde{U}_R^{(d)}$,
$\tilde{U}_L^{(u)}$, of the type \eqref{C.51} with $\alpha=0$ and $\xi_R=\xi_L$
we have
\begin{align}\label{D.22}
U_R^{(u)} &= e^{i\xi}\sigma^1\,,
\notag\\
U_L^{(u)} &= e^{i\xi}\epsilon\,,
\notag\\
\tilde{U}_R^{(d)} &= e^{i\tilde{\xi}}\sigma^1\,,
\notag\\
\tilde{U}_L^{(u)} &= e^{i\tilde{\xi}}\epsilon\,.
\end{align}
Inserting this in \eqref{D.16} we find
\begin{equation}\label{D.23}
e^{i\tilde{\xi}} = e^{i\xi}
\end{equation}
but no restriction on $\vartheta$.
We get then from \eqref{D.14}
\begin{equation}\label{D.24}
U_R^{(d)} = e^{i\xi}
  \begin{pmatrix} \sin 2\vartheta &  \cos 2\vartheta \\
                  \cos 2\vartheta & -\sin 2\vartheta \end{pmatrix}\,.
\end{equation}

For the remaining cases, $U_R^{(u)}$, $U_L^{(u)}$ according to \eqref{C.48}
and $\tilde{U}_R^{(d)}$, $\tilde{U}_L^{(u)}$ according to \eqref{C.51},
or vice versa, there is no solution of \eqref{D.16} possible.

Thus we see that, strictly speaking, the principle of maximal \CP\ invariance
plus absence of FCNCs gives no restriction on the angle $\vartheta$ in the
2--3 sector of the CKM matrix.
But perhaps we can argue that also the right-handed quarks $u_{\alpha\,R}$,
$d'_{\alpha\,R}$ should belong to some multiplet of a bigger gauge group as
would be possible in grand unified scenarios.
Then a natural requirement could be $U_R^{(u)} = U_R^{(d)}$.
From \eqref{D.19} and \eqref{D.21} as well as \eqref{D.22} and \eqref{D.24}
we see that we have then only the solution $\vartheta=0$ leading to
$V=\unitmatrix_2$.

%%%%%%%%%%%%%%%%%%%%%%%%%%%%%%%%%%%%%%%%%%%%%%%%%%%%%%%%%%%%%%%%%%%%%%%%%%%%%%
\section{\CPg\ invariances and conventional discrete symmetries}
\label{app-D}

In section~\ref{sec-discussion} we have presented our
final result for the Yukawa coupling term being compatible
with the principle of maximal \CP\ invariance; see~(\ref{5.2}).
There we have also summarised the transformations corresponding to
the invariances \CPi\ to \CPc\; see Tab.~\ref{tab-5}.
We discuss now briefly the relation of these results to conventional
discrete symmetries. 

Instead of the \CPi, \CPa, \CPb  and \CPc\ invariances
we can also consider the standard \CP\ transformation \CPb$\equiv$\CPs\
and the transformations
\begin{equation}
\label{D1}
\begin{split}
\text{D}^{(i)} &\equiv \CPi \circ \CPs,\\
\text{D}_1^{(ii)} &\equiv \CPa \circ \CPs,\\
\text{D}_3^{(ii)} &\equiv \CPc \circ \CPs,\\
\end{split}
\end{equation}
The transformations~(\ref{D1}) imply for the
first generation fermions from~(\ref{4.2}) a sign change
\begin{align}\label{D2} %and 4.3
\begin{pmatrix} \nu_{1\,L}(x) \\ l_{1\,L}(x) \end{pmatrix}
	&\rightarrow \,
    - \begin{pmatrix} {\nu}_{1\,L}(x)\\ 
                    {l}_{1\,L}(x)\end{pmatrix}\,,
\notag\\
l_{1\,R}(x)
	&\rightarrow \,
    - {l}_{1\,R}(x)\,,
\notag\\
\begin{pmatrix} u_{1\,L}(x) \\ d_{1\,L}(x) \end{pmatrix}
	&\rightarrow \,
    -\begin{pmatrix} {u}_{1\,L}(x)\\ 
                    {d}_{1\,L}(x)\end{pmatrix}\,,
\notag\\
u_{1\,R}(x)
	&\rightarrow \,
   -{u}_{1\,R}(x)\,,
\notag\\
d_{1\,R}(x)
	&\rightarrow \,
    -{d}_{1\,R}(x)\,.
\end{align}
For the Higgs fields we get, generically,
\begin{equation}
\varphi_i(x) \xrightarrow{} \widetilde{\widetilde{W}}_{ij} \varphi_j(x) ,
\end{equation}
and for the second and third fermion families
\begin{align} 
\begin{pmatrix} \nu_{\alpha\,L}(x) \\ l_{\alpha\,L}(x) \end{pmatrix}
	&\rightarrow \widetilde{\widetilde{U}}_{L\, \alpha \beta}^{(l)}   \,
     \begin{pmatrix} {\nu}_{\beta\,L}(x)\\ 
                    {l}_{\beta\,L}(x)\end{pmatrix}\,,
\notag\\
l_{\alpha\,R}(x)
	&\rightarrow \widetilde{\widetilde{U}}_{R\, \alpha \beta}^{(l)} \,
     {l}_{\beta\,R}(x)\,,
\notag\\
\begin{pmatrix} u_{\alpha\,L}(x) \\ d_{\alpha\,L}(x) \end{pmatrix}
	&\rightarrow \widetilde{\widetilde{U}}_{L\, \alpha \beta}^{(u)} \,
    \begin{pmatrix} {u}_{\beta\,L}(x)\\ 
                    {d}_{\beta\,L}(x)\end{pmatrix}\,,
\notag\\
u_{\alpha\,R}(x)
	&\rightarrow \widetilde{\widetilde{U}}_{R\, \alpha \beta}^{(u)} \,
    {u}_{\beta\,R}(x)\,,
\notag\\
d_{\alpha\,R}(x)
	&\rightarrow \widetilde{\widetilde{U}}_{R\, \alpha \beta}^{(d)} \,
    {d}_{\beta\,R}(x)\,,
\end{align}
$\alpha, \beta \in \{2,3\}$. Here the matrices $\widetilde{\widetilde{W}}$ 
and $\widetilde{\widetilde{U}}$ are given in Tab.~\ref{tab-6}

\medskip\noindent
\begin{table}
\begin{tabular}{|c|c|c|c|}
\hline
$\;\;\;\text{D}\;\;\;$ & $\;\;\;\widetilde{\widetilde{W}}\;\;\;$ & $\widetilde{\widetilde{U}}_R^{(l)}=\widetilde{\widetilde{U}}_R^{(u)}=
\widetilde{\widetilde{U}}_R^{(d)}$ & $\widetilde{\widetilde{U}}_L^{(l)}=\widetilde{\widetilde{U}}_L^{(u)}$ \\
\hline\hline
$\text{D}^{(i)}$ &$\epsilon$ & $-\epsilon$ & $-\sigma^1$\\
$\text{D}_1^{(ii)}$ &$\sigma^3$ & $\phantom{+}\sigma^3$ & $-\unitmatrix_2$\\
$\text{D}_3^{(ii)}$ &$\sigma^1$ & $\phantom{+}\sigma^1$ & $-\sigma^1$\\
\hline
\end{tabular}
\caption{\label{tab-6}
The matrices 
for the transformations of the
Higgs fields and of the second
and the third generation fermions
under the discrete symmetries $\text{D}^{(i)}$,
$\text{D}_2^{(ii)}$, and $\text{D}_3^{(ii)}$.}
\end{table}

Clearly, imposing the invariances \CPi\ to \CPc\ with the
transformations specified in Tab.~\ref{tab-5} is equivalent
to imposing \CPs\ plus the discrete invariances $\text{D}^{(i)}$ 
to $\text{D}^{(ii)}_3$ as specified in Tab.~\ref{tab-6}. But in our opinion the latter procedure
would be completely ad hoc.
We could give no physical argument for considering
just the specific transformations shown above for the Higgs fields,
the first, the second and the third fermion families. Why would we 
group the second and third families together? On the other
hand, we hope to have shown in the main text of this paper that
the principle of maximal \CP\ invariance provides us with a clear physical argument
to single out the form of the Higgs potential~(\ref{2.20}) and of
the Yukawa coupling term~(\ref{5.2}).

\end{document}